\documentclass[fleqn,usenatbib]{mnras}

\usepackage{newtxtext,newtxmath}
\usepackage[T1]{fontenc}

\DeclareRobustCommand{\VAN}[3]{#2}
\let\VANthebibliography\thebibliography
\def\thebibliography{\DeclareRobustCommand{\VAN}[3]{##3}\VANthebibliography}

\newcommand\dd{{\rm d}}
\newcommand{\cs}{c_{\rm s}}
\def\Omegap{\Omega_{\rm p}}
\def\pa{\partial}
\newcommand{\bfv}{\mathbf{v}}
\newcommand{\kms}{\,{\rm km\, s^{-1}}}
\newcommand{\kpc}{\,{\rm kpc}}
\newcommand{\Myr}{\,{\rm Myr}}

\usepackage{graphicx}	% Including figure files
\usepackage{amsmath}	% Advanced maths commands
\usepackage{soul}
\usepackage{hyperref} 

\title[Tremaine-Weinberg method using gas tracers]{On the Tremaine-Weinberg method: how much can we trust gas tracers to measure pattern speeds?}

\author[O. Borodina et al.]{
Olga Borodina,$^{1, 2}$\thanks{E-mail: olga.borodina@cfa.harvard.edu}
Thomas G. Williams,$^{3, 1}$
Mattia C. Sormani,$^{4}$
Sharon Meidt,$^{5}$
and Eva Schinnerer$^{1}$
\\
$^{1}$Max Planck Institut f{\"u}r Astronomie, K{\"o}nigstuhl 17, 69117 Heidelberg, Germany\\
$^{2}$Center for Astrophysics $\vert$ Harvard \& Smithsonian, 6P Garden St, Cambridge, MA 02138, USA\\
$^{3}$Sub-department of Astrophysics, Department of Physics, University of Oxford, Keble Road, Oxford OX1 3RH, UK\\
$^{4}$Universit{\"a}t Heidelberg, Zentrum f{\"u}r Astronomie, Institut f{\"u}r theoretische Astrophysik, Albert-Ueberle-Str. 2, 69120 Heidelberg, Germany \\
$^{5}$Sterrenkundig Observatorium, Universiteit Gent, Krijgslaan 281 S9, B-9000 Gent, Belgium
}

\date{Accepted XXX. Received YYY; in original form ZZZ}

\pubyear{2022}

\begin{document}
\label{firstpage}
\pagerange{\pageref{firstpage}--\pageref{lastpage}}
\maketitle

\begin{abstract}
Pattern speeds are a fundamental parameter of the dynamical features (e.g. bars, spiral arms) of a galaxy, setting resonance locations. 
Pattern speeds are not directly observable, so the Tremaine-Weinberg (TW) method has become the most common method used to measure them in galaxies. However, it has not been tested properly whether this method can straightforwardly be applied to gas tracers, despite this being widely done in the literature. When applied to observations, the TW method may return invalid results, which are difficult to diagnose due to a lack of ground truth for comparison.
Although some works applying the TW method to simulated galaxies exist, only stellar populations have been tested. Therefore, here we explore the applicability of the TW method for gas gracers, by applying it to hydrodynamical simulations of galaxies, where we know the true value of the bar pattern speed.
We perform some simple tests to see if the TW method has a physically reasonable output. We add different kinds of uncertainties (e.g. in position angle or flux) to the data to mock observational errors based on the magnitude of uncertainty present in the observations. Second, we test the method on 3D simulations with chemical networks. We show that in general, applying TW to observations of gas will not recover the true pattern speed. These results have implications for many ``pattern speeds'' reported in the literature, and based on these tests we also give some best practices for measuring pattern speeds using gas tracers going forwards.
\end{abstract}


%%%%%%%%%%%%%%%%%%%%%%%%%%%%%%%%%%%%%%%%%%%%%%%%%%

%%%%%%%%%%%%%%%%% BODY OF PAPER %%%%%%%%%%%%%%%%%%

\section{Introduction}

In the local Universe, 30\% to 50\% of galaxies are barred \citep{Sheth2008, BinneyTremaine2008, Aguerri2009}. 
Bars are believed to rotate with a well-defined pattern speed which is one of the most important parameters because it sets the location of the corotation and Lindblad resonances.
Bars can also have a profound impact on galaxy evolution, causing starburst events at the interface with spiral arms \citep{Beuther2017}  and can lead to star formation suppression along the bar \citep{Querejeta2021}. Furthermore, pattern speeds are a possible key to understand the interaction between the bar and dark matter halo \citep{Hernquist1992,Debattista2000, Weinberg2007, Beane2022}. Therefore, the accurate measurement of pattern speeds is vital to understanding large-scale dynamical structures in galaxies.

Pattern speeds can not be observed directly, but there are several different ways to measure them. For example, we can estimate the pattern speed from the velocity at radii that correspond to resonances in the disc \citep{Elmegreen1989, Kuno2000}, or we can match a simulation, where the pattern speed is already known, to observed galaxies \citep{Weiner2001, Hirota2009, Lin2013, Sormani2015}. Galaxy modelling has been employed to robustly define the pattern speed, and it can produce smaller uncertainties in the measurements \citep{Hunter1988, Sempere1995, Salo1999, Rautiainen2008, Kalapotharakos2010}. However, the physical link between observational features (e.g. bar ends) and dynamical features, like corotation, is still not firmly established or expected in all cases \citep{Kranz2003, Williams2021}. Plus, running suites of bespoke simulations is computationally expensive, limiting the usefulness of these methods.

\cite{TW1984} developed a model-independent method to calculate pattern speeds. It has become favored due to its apparent simplicity, requiring only line-of-sight velocity and brightness information along the direction parallel to the galaxy major axis
In most modern applications one can get the pattern speed even in a single observation with interferometric imaging (e.g., HI, CO with VLA and ALMA respectively), Fabry-Perot observations \citep{Debattista2004, Chemin2009} and more recently using wide field of view optical integral-field unit (IFU) spectroscopy \citep{Guo2019, Cuomo2019, Cuomo2020} (e.g., stars or $\text{H}{\alpha}$ with MUSE).
During the last decades the TW method has been applied to many different tracers such as stars \citep{Gerssen2003, Corsini2007, Cuomo2020, Buttitta2022} or gas: H{\sc i} \citep{Bureau1999, Banerjee2013}, CO \citep{Zimmer2004}, and $\text{H}{\alpha}$\citep{Chemin2009}. 
At first sight using gas as a tracer is preferable as its emission lines are bright and easier to study rather than composite stellar spectra, which require complex modelling. Therefore, the application of the TW method to CO and H{\sc i} has regularly been used over the past three decades to measure pattern speeds, as it is usually assumed that stars and gas have the same pattern speed \citep{Sellwood1993}. Given the purely data-driven nature of the TW method, it is also straightforward to apply to simulations. For example,  \cite{Roshan2021} have used the TW method to calculate pattern speeds from single snapshots in the IllustrisTNG \citep{Nelson2018, Pillepich2019} and EAGLE simulations \citep{Schaye2015}. 

However, using data from the Physics at High Angular resolution in Nearby GalaxieS (PHANGS) survey \cite{Williams2021} showed that when we apply the TW method to different tracers it can yield different ``pattern speeds'', indicating that different tracers (e.g. stars or ionised/molecular gas) may be compromised in different ways. Therefore, we need to carefully test this method on mock data. For instance, we can apply the TW method to $N$-body simulations \citep{Debattista2003,Gerssen2007, Zou2019, Guo2019} to study the limitations of the method by comparing the output of the method with the pattern speed that was set by the model (a ground truth or GT). These works found that the inclination range and bar position alignment with the position angle of the galaxy on the sky should be restricted (the reasons for this will be discussed later in sect. \ref{sect:twm}). It is also clear from these simple simulations that the method is extremely sensitive to position angle (PA) measurements \citep{Debattista2003}. 

Despite previous work, it is still not fully understood to what extent the TW method can be applied to gas tracers. The biggest concern is that the gas does not obey the continuity equation which is one of the fundamental assumptions of the TW method (see sect. \ref{sect:twm}). This is caused by the baryon life cycle, i.e. atomic gas is converted into molecular gas, then star formation processes molecular clouds into stars that ionize the gas \citep{ Schinnerer2019Msngr, Schinnerer2019}.
Furthermore, this method has never been tested properly on 3D hydrodynamical simulations. Therefore, here we revisit the question of the applicability of the TW method to gas tracers.

%Paper structure
The structure of this paper is as follows: we first describe the simulations we use and how we create mock observational data (Section~\ref{sect:method}). Then we briefly explain the method itself. In Section~\ref{sect:results} we present the main results and then describe their implications in Section~\ref{sect:discussion}. Finally, we summarize the conclusions in Section~\ref{sect:conclusions}.

\section{Method}
\label{sect:method}

\subsection{The simulations}

In this work, we have used two different kinds of simulations. First, we use a simple 2D isothermal simulation of gas flow in an external bar potential. These simulations have a smooth gas density distribution and allow us to test the TW method in the simplest possible setup. Then, we use the 3D simulation presented in \cite{Sormani2018} which includes non-equilibrium chemical networks. The external gravitational potential of the bar is the same as in the first simulation. These 3D simulations have a clumpy interstellar medium (ISM), and the continuity equation does not apply to individual tracers such as CO and H{\sc i} which trace the molecular and atomic gas phases. These more complex simulations, therefore, allow us to assess how the TW method performs when some of the underlying assumption of the method (smoothness and absence of sources and sinks) are not perfectly satisfied.

\subsubsection{Numerical setup of the 2D simulations} \label{sect:sims2D}

We ran 2D isothermal non-selfgravitating hydrodynamical simulations in an externally imposed, rotating barred potential. We use the public grid code {\sc Pluto} \citep{Mignone2007} version 4.3. The external gravitational potential is exactly the same as described in Section 3.2 of \cite{Ridley2017}, and the simulations here are very similar to those described in that paper except for the different code and grid geometry used (cartesian by \citealt{Ridley2017} vs polar here for a better resolution in the centre). The gravitational potential is constructed to reproduce the properties of the Milky Way, which serves here as a template for a barred galaxy. The exact same potential is also used in the 3D simulations described below from \cite{Sormani2018}. The potential is assumed to be rigidly rotating with bar pattern speed of $\Omegap=40 \rm \, km \, s^{-1} \, kpc^{-1}$.

We assume that the gas is isothermal, i.e.
\begin{equation}
P = \cs^2 \Sigma \,,
\end{equation}
where the sound speed is $\cs = 10 \kms$ and $\Sigma$ describes the gas surface density. We neglect the gas self-gravity and its associated additional physics. The equations of motion in the rotating frame co-rotating with the bar are the continuity and Euler equations:
\begin{align}
\pa_t \Sigma + \nabla\cdot \left( \Sigma \bfv \right) &= 0 \label{eq:continuity} \\
\pa_t \bfv + \left(\bfv \cdot \nabla \right) \bfv &= - \cs^2 \frac{\nabla \Sigma}{\Sigma} - \nabla\Phi - 2 {\Omegap} \hat{\mathbf{e}}_z \times \bfv + \Omegap^2 R \, \hat{\mathbf{e}}_R \label{eq:euler}
\end{align}
where $\Sigma$ is the surface density, $\bfv$ is the velocity, $(R,\theta,z)$ denote standard cylindrical coordinates, $\hat{\mathbf{e}}_R$ is the unit vector in the radial direction and $\hat{\mathbf{e}}_z$ the unit vector in the $z$ direction.

We use a two-dimensional static polar grid covering the region $R \times \theta = [0.01,10] \kpc \times [0, 2\pi]$. The grid is logarithmically spaced in $R$ and uniformly spaced in $\theta$ with $1024 \times 1024$ cells. We use the following parameters: {\sc rk2} time-stepping, no dimensional splitting, {\sc hll} Riemann solver and the default flux limiter. We solve the equations in the frame rotating at $\Omegap$ by using the {\sc rotating\_frame = yes} switch. Boundary conditions are reflective on the inner boundary at $R=0.01\kpc$ and outflow on the outer boundary at $R=10.0\kpc$.

\begin{figure}
    \centering
    \includegraphics[width=\columnwidth]{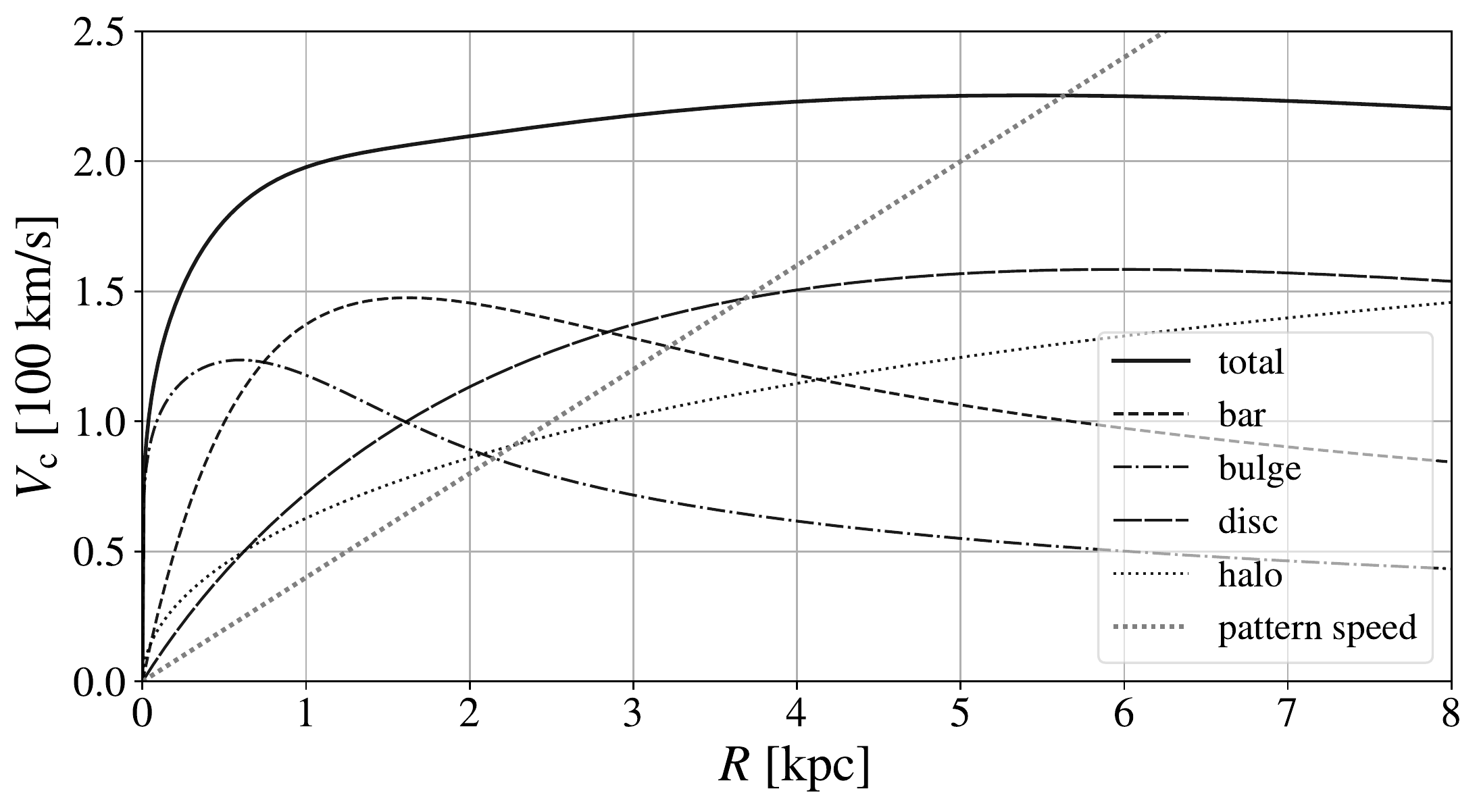}
    \caption{Rotation curves for each component in our simulation potential. The bar is rigidly rotating with a pattern speed of $\Omega_{\rm p} = 40 \text{km} \, \text{s}^{-1} \, \text{kpc}^{-1}$}.
    \label{fig:vcirc}
\end{figure}

The initial density distribution is taken to be
\begin{equation}
    \Sigma_0 = A \exp\left( -\frac{R_m}{R} -\frac{R}{R_d} \right)
\end{equation}
where $R_m=1.5\kpc$, $R_d=7\kpc$ and without loss of generality (the equations of motion are invariant for density rescaling, so the density units are arbitrary) we set $A=1$. In order to avoid transients, we introduce the bar gradually, as is common practice in this type of simulations \citep[e.g.,][]{Athanassoula1992}. We start with gas in equilibrium on circular orbits in an axisymmetrised potential and then we linearly turn on the non-axisymmetric part of the potential during the first $150\Myr$ while keeping the rotation curve fixed (Fig. \ref{fig:vcirc}).

\subsubsection{Numerical setup of the 3D simulations}
\label{sect:sims3D}

The simulation used here is the ``variable'' simulation from \cite{Sormani2018}. We give here only a very brief overview, and refer to that paper for a more complete description.

The simulations are run using the moving-mesh code {\sc arepo} \citep{Springel2010,Weinberger2020}. They are three-dimensional and unmagnetised, and include a live chemical network that keeps track of hydrogen and carbon chemistry. In particular, for the purposes of this paper it is important that we can calculate the amount of molecular CO and atomic H{\sc i} at each $(x,y,z)$ point. Gas self-gravity and star formation are neglected. The simulations comprise interstellar gas out to a galactocentric radius of $R \leq 8 \kpc$.

\subsection{Simulation post-processing}
\label{sect:postproc}

\begin{figure*} % it's here for layout
    \centering
    \includegraphics[width=0.9 \textwidth]{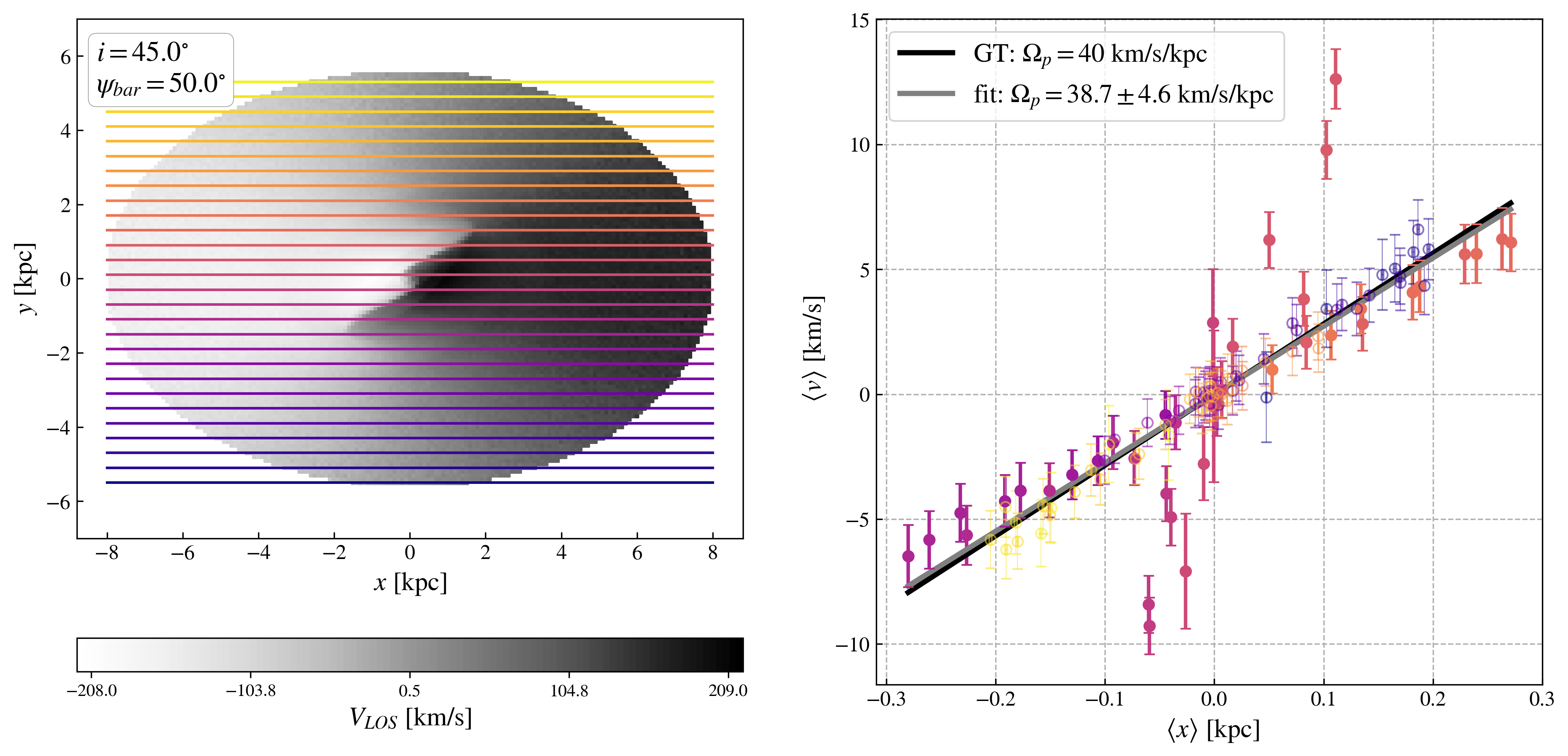}
    \caption{Applying the TW method to our simulated 2D galaxy. \textit{Left:} grey-scale map shows the line-of-sight velocity field and horizontal lines indicate the centres of our 100~pc slits. Only one in every four slits is shown, for legibility. \textit{Right:} $\langle v \rangle$ and $\langle x \rangle$, each data point corresponds to the matching colour slit in the left panel. The black line shows the ground truth (GT) pattern speed and the grey line is the fit to slits that cross the bar that are shown as filled circles. Slits which are not included in the fitting are marked with open circles.}
    \label{fig:strongbar}
\end{figure*}

From the simulations, we create a mock pixelated image to mimic gas observations.
Initially, we have the bar aligned with the $x$-axis, so we rotate the galaxy by angle $\psi_{\text{bar}}$ around the $z$-axis. Second, we incline the galaxy along the $x$-axis (which now does not in general coincide with the major axis of the bar) by angle $i$.
We calculate the line-of-sight velocity as follows:
\begin{equation}
    v_{\text{LOS}} (x, y) = v_y (x, y, z) \sin(i) + v_z (x, y, z) \cos(i) \, ,
\end{equation}
where $v_z (x, y, z) \equiv 0$ for the 2D simulation.
For 3D simulations, we also weight velocity by the mass of particles in the bin.

To make these simulations appear closer to observational data we add Gaussian noise with a standard deviation of 10 km~s$^{-1}$ for the velocity field and 5\% uncertainty for the density of each pixel.
In real observational data,  we do not know the exact location of the galaxy centre. Therefore, we add a centering error, in which values are picked from a Gaussian distribution with a standard deviation equal to the slit width $h(y) = 100$\,pc.
To mock uncertainty in position angle (PA) measurements, we add another rotation of the inclined galaxy by $\delta_\text{PA}$ angle, which is drawn from a normal distribution with a standard deviation of $1^\circ$. Error values estimations for flux, velocity, and PA were based on typical values from PHANGS (Physics at High Angular resolution in Nearby GalaxieS) survey ALMA data \citep{Leroy2021}, which represent the current smallest uncertainties achievable with relatively large surveys.

\subsection{Tremaine-Weinberg method}
\label{sect:twm}
Our work is based on the formula presented by \cite{TW1984}:
% \begin{equation}
\begin{align}
    \Omega_\text{P} \sin(i) = \frac{\int\limits_{-\infty}^{\infty}h(y)\int\limits_{-\infty}^{\infty}v_{\text{LOS}} (x, y) \Sigma (x, y) \dd x \dd  y}{\int\limits_{-\infty}^{\infty}h(y)\int\limits_{-\infty}^{\infty}\Sigma (x, y) x \dd x \dd y} = \frac{\langle v \rangle}{\langle x \rangle} \, ,
\label{eq:TWM}
\end{align}
% \end{equation}
where $h(y)$ is the weight function, which in our case has a form of a boxcar function to represent a slit, mimicking long-slit spectroscopy, or columns of pixels in IFU data.

This formula is based on the following three assumptions \citep{TW1984}:
\begin{enumerate}
    \item The disc of the galaxy is flat.
    \item The disc has a single well-defined and constant pattern speed
    \item The tracer should obey the continuity equation, i.e.\ it has neither sources nor sinks.
\end{enumerate}
Because this method is designed to catch non-axisymmetric structure, any deviation from axisymmetry is assumed to be caused by the pattern.

From the formula we can see that the method should be applied to moderately-inclined galaxies. For edge-on galaxies we will be unable to identify the bar, and for face-on galaxies the line-of-sight velocity is too small. The same logic can be used for bar alignment restrictions: when the bar is oriented along either the major or minor kinematic axis of a galaxy, no left-right asymmetry is present and the integral will evaluate to zero.

We  bin the simulation into 100\,pc-sized `pixels'. 
Due to computational asymmetry or rotation by $\delta_\text{PA}$ and centre shifting, the number of pixels on either side of the galaxy centre may not be exactly equal (i.e. $N(x<0) \neq N(x>0)$).
Therefore, we symmetrize both $\Sigma (x, y)$ and $V_{\text{LOS}} (x, y)$ by setting pixels without a corresponding opposite counterpart to zero, so $N(x<0) = N(x>0)$ along each slit to minimise non-axisymmetries induced simply from the pixelisation process

Then we calculate $\langle v \rangle$ and $\langle x \rangle$ and fit data points $(\langle x \rangle$, $\langle v \rangle)$ for those slits which cross the bar using orthogonal distance regression \citep[ODR, ][]{scipy2020}. The slope of $(\langle x \rangle$, $\langle v \rangle)$ is then simply the pattern speed $\Omega_\text{P} \sin i$.

We include uncertainties in density $\Sigma (x, y)$ and velocity $v_{\text{LOS}} (x, y)$ by adding noise values sampled from a Gaussian distributions with standard deviation of 5\% of pixel's density and 10 $\text{km~s}^{-1}$, respectively. However, different values of noise in data and $\delta_\text{PA}$ lead to different pattern speed measurements. Therefore, we implement a bootstrapping procedure to estimate the uncertainty in our fits.
We repeat the measurements 500 times, each time  randomly sampling from our noise distribution.
Then we calculate the median and 16$^{\rm th}$ and 84$^{\rm th}$ percentiles to define pattern speed error bars. We will use these values as the nominal pattern speed and its uncertainties.

\section{Results} 
\label{sect:results}

\subsection{Hydrodynamical 2D simulations}

We first applied the TW method to the simple 2D hydrodynamical simulation of the galaxy. 
This simulation obeys the first and third assumptions \citep{TW1984}, i.e. the disc is flat and gas obeys the continuity equation by design. However, the gas never reaches a perfect steady state while flowing in the bar potential. Therefore, the density distribution changes slightly from snapshot to snapshot in a frame rotating with the pattern speed $\Omega_\text{P} = 40 \text{km} \, \text{s}^{-1} \, \text{kpc}^{-1}$. We have checked that repeating the analysis for other time snapshots does not change the conclusions in this Section.

As shown in Figure \ref{fig:strongbar}, we recover the correct pattern speed  $38.7 \pm 4.6 \, \text{km} \, \text{s}^{-1} \, \text{kpc}^{-1}$.
However, we can also see that there is a second steep slope which corresponds to the slits crossing the center of the bar. We discuss it further in Section \ref{sect:discussion}. Moreover, the slits outside the bar, i.e. where there is no pattern (yellow and blue color in the Fig. \ref{fig:strongbar}), have non-zero  $\langle v \rangle$ and $\langle x \rangle$ and these points also follow the ground truth line.

Intriguingly, we appear to be measuring a pattern speed even in the outer region of the disc where the density perturbations induced by the bar are expected to be negligibly small. Similarly, recent results from \cite{Williams2021} show that the use of a gas tracer can lead to erroneous pattern speed measurements. To test whether these measurements are real or simply an artifact in the data, we take a step back and perform a simple test with the disc with solid body rotation.

\subsection{Semi-analytical 2D simulations}

To test when the TW method can pick up errant signal, we performed a very simple but unphysical test. We created a mock galaxy using an exponential 2D density profile, which rotates as a solid body with an angular speed of 40 $\text{km} \, \text{s}^{-1} \, \text{kpc}^{-1}$, using the positions of particles from the previously described 2D simulation as a basis. We will refer this as our `perfect disc model'. As before, we also added noise to the data. Due to absence of any left-right asymmetry, we would expect to have zero values for both $\langle v \rangle$ and $\langle x \rangle$ because signals for $x < 0$ and $x > 0$ under the integral sign will cancel out. As we will show, this is the case but only under certain fairly strict conditions. Also, we stress that this model does not reflect reality, but we use it to illustrate the conditions where we can pick up false signal using the TW method in a case where we should not detect any signal.
 
 \subsubsection{Effect of data uncertainties}
 \label{sect:perfectdisk}
 
\begin{figure*}
    \centering
    \includegraphics[width=0.9 \textwidth]{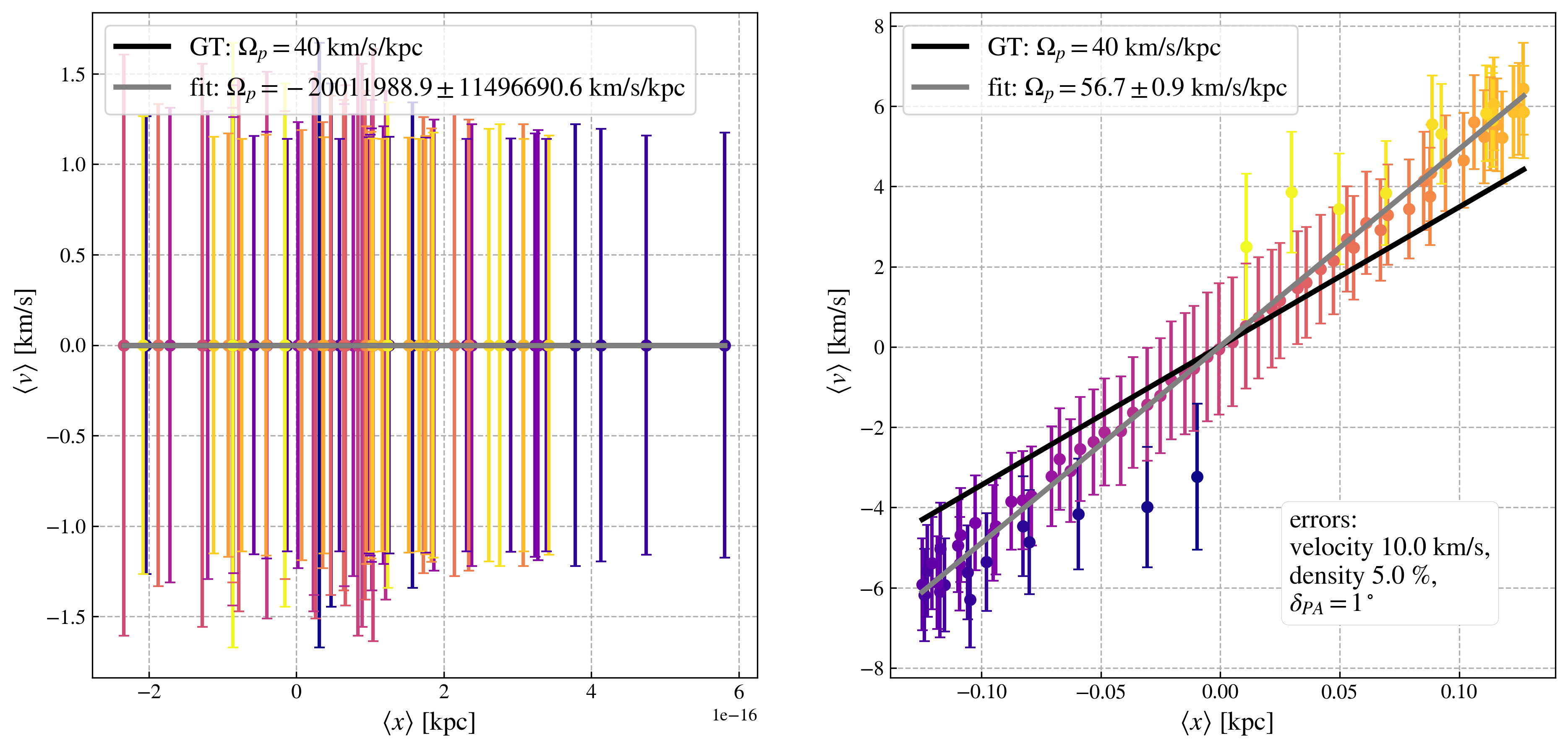}
    \caption{Applying the TW method to our perfect disc model with $\psi_{\text{bar}} = 90^\circ$ and $i = 45^\circ$. \textit{Left:} $\langle v \rangle$ and $\langle x \rangle$ plot for a galaxy with $\delta_\text{PA} = 0^\circ$, and without noise in densities and velocities. We can see that no signal is detected here, and the fitted pattern speed is consistent with 0~km~s$^{-1}$~kpc$^{-1}$. \textit{Right:} $\langle v \rangle$ and $\langle x \rangle$ plot for a galaxy with $\delta_\text{PA} = 1^\circ$ and noise added to the density and velocity data. The black line here shows the solid body rotation of the perfect disc and the grey line is the fit. In this case, even though no pattern is present the TW method detects signal close to the solid body rotation of the disk.}
    \label{fig:perfectdisk}
\end{figure*}

\begin{figure}
    \centering
    \includegraphics[width=\columnwidth]{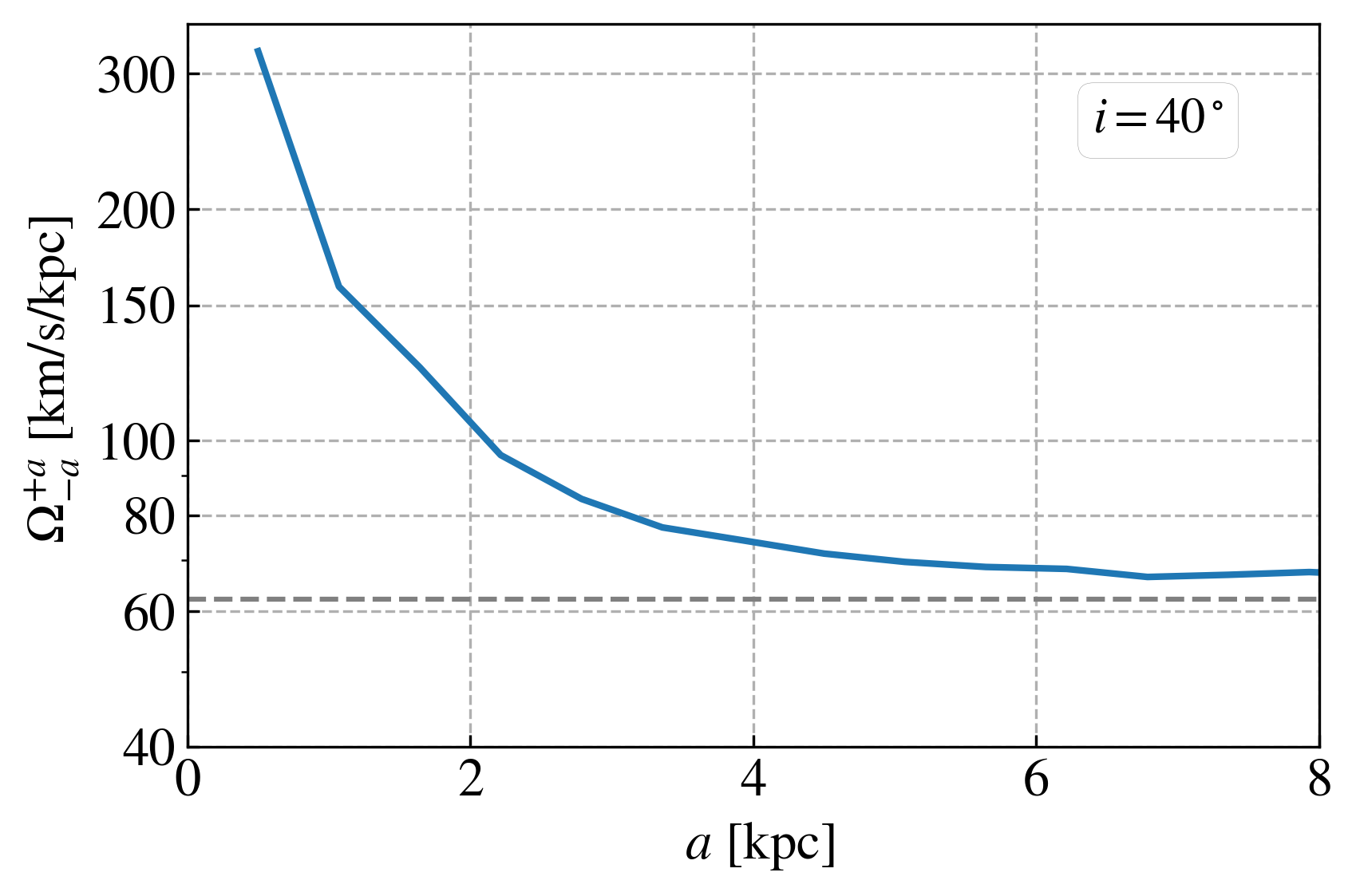}
    \caption{Blue line shows measured pattern speed for slits integrated between $-a$ to $a$ for the perfect disc model with an inclination of $i = 50^\circ$. It does not reach SBR value $\Omega = 40$ km~s$^{-1}$~kpc$^{-1}$ due to reasons discussed below. Gray dashed line corresponds to the value of  $\Omega / \sin(i)$}
    \label{fig:edges}
\end{figure}

 Firstly, we confirmed that if we have no uncertainties in the data (i.e. no noise in the density distribution, velocity, or in PA), we do not measure any significant signal (left panel in Fig. \ref{fig:perfectdisk}). However, if we include either density uncertainties or PA error then we measure a non-zero pattern speed (right panel in Fig. \ref{fig:perfectdisk}). We emphasize that symmetrization of pixels, which was described above, does not help to cancel out the signal because it is not an edge effect, the asymmetry appears along the whole slit (see Fig. 1 in \citealt{GarmaOehmichen2020}). We demonstrate this effect in Figure \ref{fig:edges} where we calculated the pattern speed for the perfect disc model with slits that do not cross the whole galaxy but start from $x = -a$ and end at $x = a$. For higher inclinations we see the pattern speed tends towards the value for the solid body rotation.

Nevertheless, adding 5\% uncertainty to the density values is enough to create a false signal of the pattern speed. It happens because noise in the data is like having "clumps" and "holes". When $\Sigma (-x) \neq \Sigma (x)$ pixels do not cancel out under the integral, we see a non-zero pattern speed measurement within that slit. We can illustrate this with the simplest example: if there is only one pixel with a clump, then the TW method measures the rotation of that clump. Therefore, we measure a rotation velocity weighted by $\overline{\Sigma (x) x}$ (from Eq. \ref{eq:TWM}) with those clumps.

\begin{figure}
    \centering
    \includegraphics[width=\columnwidth]{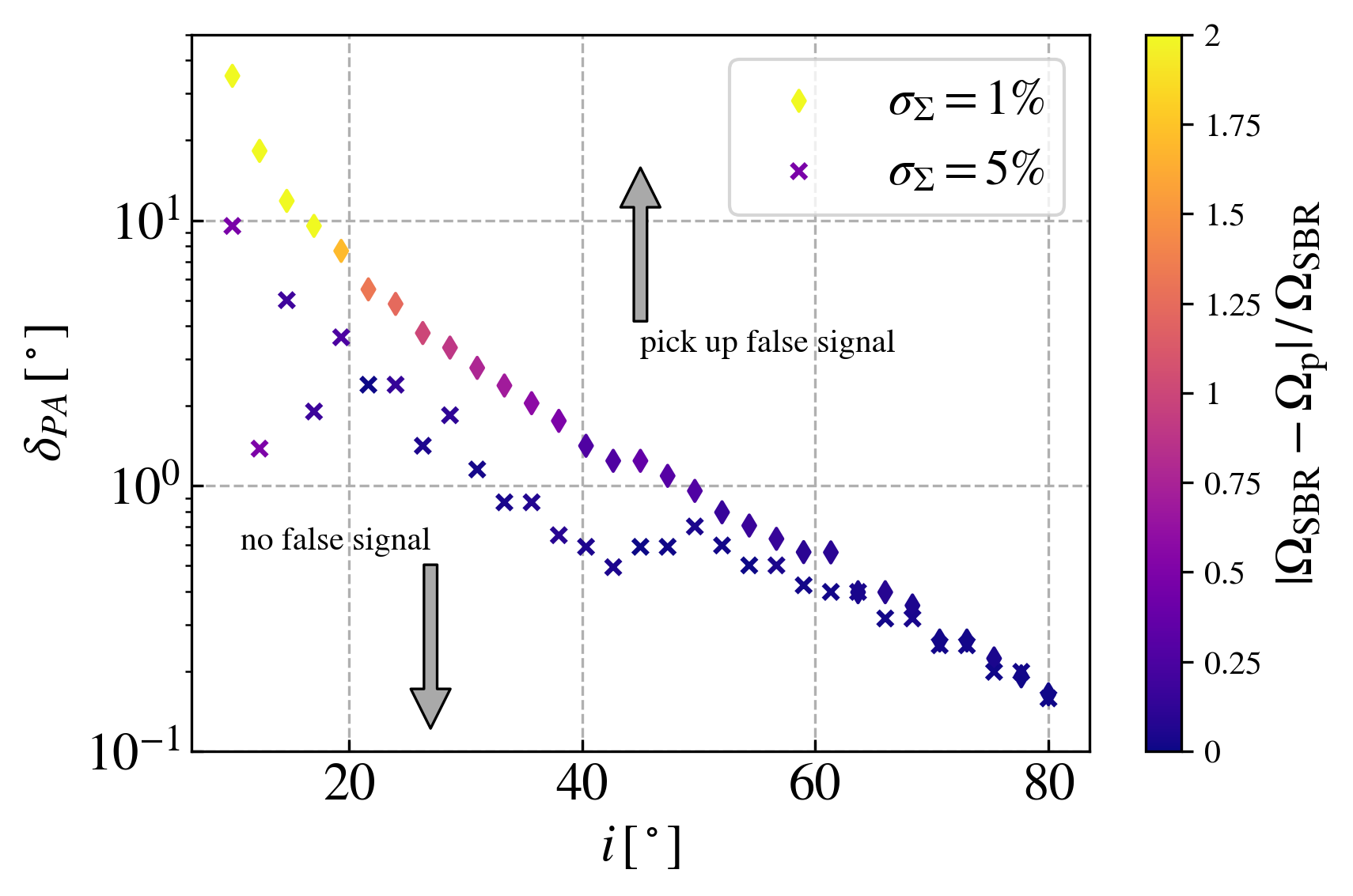}
    \caption{Allowable $\delta_\text{PA}$ uncertainty for different inclinations and density noise levels. If the PA uncertainty is below diamonds or crosses, then we do not pick up any false signal. Diamonds correspond to flux uncertainty of 1\% and crosses correspond to an uncertainty of 5\%. For lower inclinations, higher than $\delta_\text{PA}>1^\circ$ are acceptable. The colour of the points shows the difference between the fitted $\Omega_\text{P}$ and the GT.}
    \label{fig:deltaPA_incl}
\end{figure}

Thus, we want to find the minimal value of $\delta_\text{PA}$ corresponding to the fixed inclination where the TW method will start picking up signal that is not due to a pattern. 
We applied the TW method to the perfect disc model for the range of inclinations $ 10^\circ < i < 80^ \circ$ using the bootstrap procedure for different uncertainties in PA and centering. Due to the fact that this model does not have any pattern we would expect some noise instead of the pattern speed. However, uncertainties in data lead the TW method to pick up a signal. In Figure \ref{fig:deltaPA_incl} we plot these values of $\delta_\text{PA}$, which correspond to the $\delta_\text{PA}$ values where the TW method produces a non-zero $\Omega_\text{P}$. The relative difference between measured ``pattern speed'' and solid body rotation (SBR) velocity is highlighted by the colour of the point. This shows that for $i<20^\circ$, the error in PA can be up to $1^\circ$ without detecting a false signal. Also, in Figure \ref{fig:deltaPA_incl} we show that the level of flux uncertainty has an impact on these values. Higher values of noise result in more asymmetry, and make the constraint on position angle uncertainty more stringent.

\subsection{The effect of bar alignment and galaxy inclination}
\label{sect:i-psibar-2D}

As well as limitations on $\delta_\text{PA}$, we also expect to have limitations for bar alignment and galaxy inclinations where the TW method will break down. For example, if the galaxy is too inclined, we might not have enough slits to fit for the pattern speed (which may well be a function of resolution) and also as shown in Figure \ref{fig:deltaPA_incl} we can catch average rotation velocity. Given that the bar is a roughly straight structure, there may also be regimes where the bar is sufficiently aligned or perpendicular to the galaxy's kinematic major axis that we are unable to pick up any signal in the integrals.
Therefore, we applied the TW method to galaxies with a range of inclinations $2^\circ < i < 88^\circ$ and a range of bar alignment angle $0^\circ < \psi_{\text{bar}} < 180^\circ$ using the bootstrapping procedure described above (see end of Sect. \ref{sect:postproc}) to test the limitations imposed by bar (mis-)alignment and galaxy inclination.

\begin{figure}
    \centering
    \includegraphics[width=0.7\columnwidth]{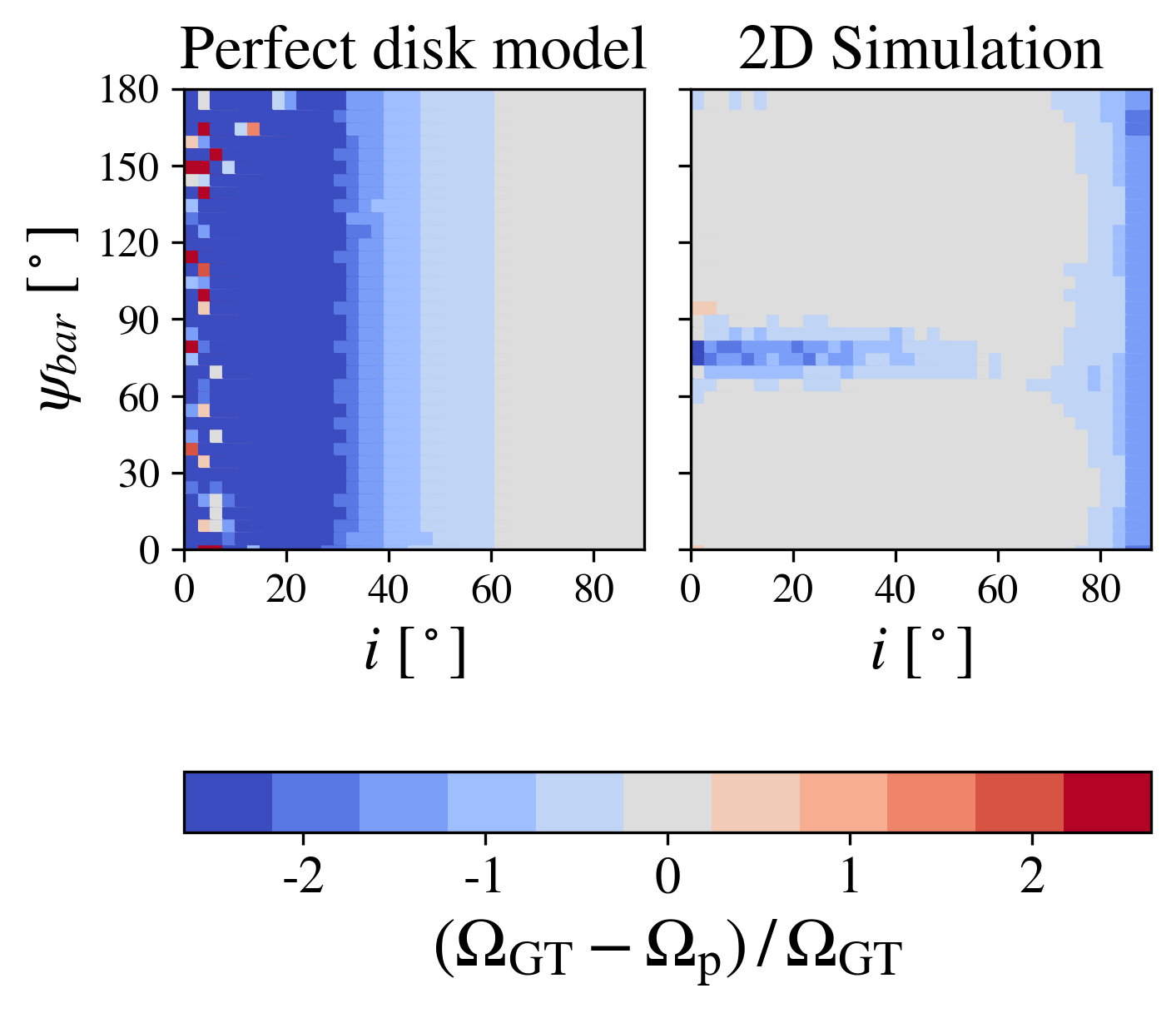}
    \caption{The bootstrapped TW method output for a range of angles $i$ and $\psi_{\text{bar}}$ for 2D test models. Color shows the relative difference between ground truth pattern speed ($\Omega_\text{GT}$) and measured pattern speed ($\Omega_\text{P}$) -- gray indicates agreement and increasing blue (red) shows increasing over- (under-) prediction.
    \textit{Left:} Results for perfect disc test, with no bar and hence no pattern speed. For small inclinations pattern speed measurements are random.
    \textit{Right:} Results for 2D simulations, including a bar. 
    }
    \label{fig:bootstrap}
\end{figure}

In Figure \ref{fig:bootstrap}, we see how the TW method outputs differ depending on $i$ and $\psi_{\text{bar}}$ for two simple models -- the regular 2D simulation, described in Section \ref{sect:sims2D} and the perfect disc (Sect. \ref{sect:perfectdisk}). Firstly, for a perfect disc with no pattern speed (see left panel of Fig. \ref{fig:bootstrap}) we see that for small inclinations we measure noise as expected (see Fig. \ref{fig:perfectdisk}). However, the more inclined galaxy we have, the more likely we catch the solid body rotation instead of pattern speed.
Second, we see that bar orientation matters (right: 2D hydrodynamical simulation). When the bar is located along the minor or major axis of galactic projection we do not recover the true pattern speed. However, due to adding uncertainties in the PA and centering, this introduces some asymmetry, and the blue region on the right panel of Fig. \ref{fig:bootstrap} is shifted down from being centred on $\psi_{\text{bar}} = 90^\circ$. If we do not include these uncertainties, 
this blue region is shifted back up to be centred at $\psi_{\text{bar}} = 90^\circ$.
this blue region is shifted back up to be centred at $\psi_{\text{bar}} = 90^\circ$.

We define a lower bound on the inclination of $5^\circ$, which is highly dependent on the measured uncertainties of the velocities and densities. For this lower bound, we assume an uncertainty (per pixel) of 10 $\rm km~s^{-1}$ and 5\%, respectively.

The upper limit for inclinations we define as follows. We measure some signal at very low inclinations, but the difference between the measured speed and the true pattern speed is greater than a factor of two (left panel of Fig. \ref{fig:bootstrap}). It would give us  an error in corotation radius more than 50\%, so it is easy to detect this discrepancy and remove questionable slits from consideration. However, for inclinations larger than $i_\text{max}$ we might consider a false signal as the real one. Therefore,
from Figure \ref{fig:bootstrap} we see that $i_\text{max} = 50^\circ$ and we should not apply the TW method to galaxies with higher inclinations.

In summary, we have used 2D simulations to highlight the effects of a number of parameters, and provide limits on various uncertainties and geometric parameters in obtaining reliable results from the TW method. Firstly, we suggest a lower bound to the galaxy inclination of $5^\circ$ and an upper bound of $i =  50^{\circ}$. These conclusions are based on unphysical model of a galaxy with the properties described above. Secondly, bar alignment matters too and reasonable results are obtained for $5^{\circ} < |\psi_{\text{bar}}| < 30^{\circ}$. We caution that even when these conditions are met, and with uncertainties matched to the best currently available data, given the inherent noise in the observations that false signal may still be picked up by the TW method.

\subsection{Hydrodynamical 3D simulations}

Given that observationally we do not have access to the total gas density but only to approximate gas tracers \citep{Bureau1999, Zimmer2004, Chemin2009,  Banerjee2013}, we also wish to explore how the choice of tracer effects the results of the TW method. We use the 3D hydrodynamical simulations of \cite{Sormani2018} that include a non-equilibrium chemical network that captures the multi-phase nature of the ISM. These simulations are well-suited to our present purpose, as they can be used to predict maps of CO and H{\sc i} emission (see density maps in Appendix \ref{sect:densitymaps}), although they do not include star formation, so we will not explore the effects of using $\text{H}{\alpha}$ in this study. As earlier for 2D simulations, we did the same post-processing to the data, e.g. rotation and adding noise and imperfections, and additionally we produce a 2D image from this 3D simulation using an assumed inclination angle. 

First, we used the total gas content to check if the TW method measures the correct pattern speed for this model, which we would expect to be the case as given the lack of star formation the continuity equation will hold. Second, we applied the TW method to individual gas tracers, such as CO or H{\sc i} only, as they are commonly used observationally. Third, we constructed a mock hydrogen tracer. As the cold H$_2$ is impossible to observe directly, we usually use CO as a molecular gas indicator. Therefore, we calculated H$_2$ density as if we did not have it from the simulation:
\begin{equation}
      \Sigma = \Sigma_\text{H{\sc i}} + \Sigma_\text{CO} \cdot \frac{\left< \Sigma_{\text{H}_2}\right>}{\left<\Sigma_\text{CO}\right>}  \,,
\end{equation}
where $<\Sigma_{\text{H}_2}>$ and $<\Sigma_\text{CO}>$ are the median values for the H$_2$ and CO surface densities. This is an equivalent of a single ``CO-conversion factor'', $\alpha_{\rm CO}$, in observational work.

\begin{figure*}
    \centering
    \includegraphics[width=14cm]{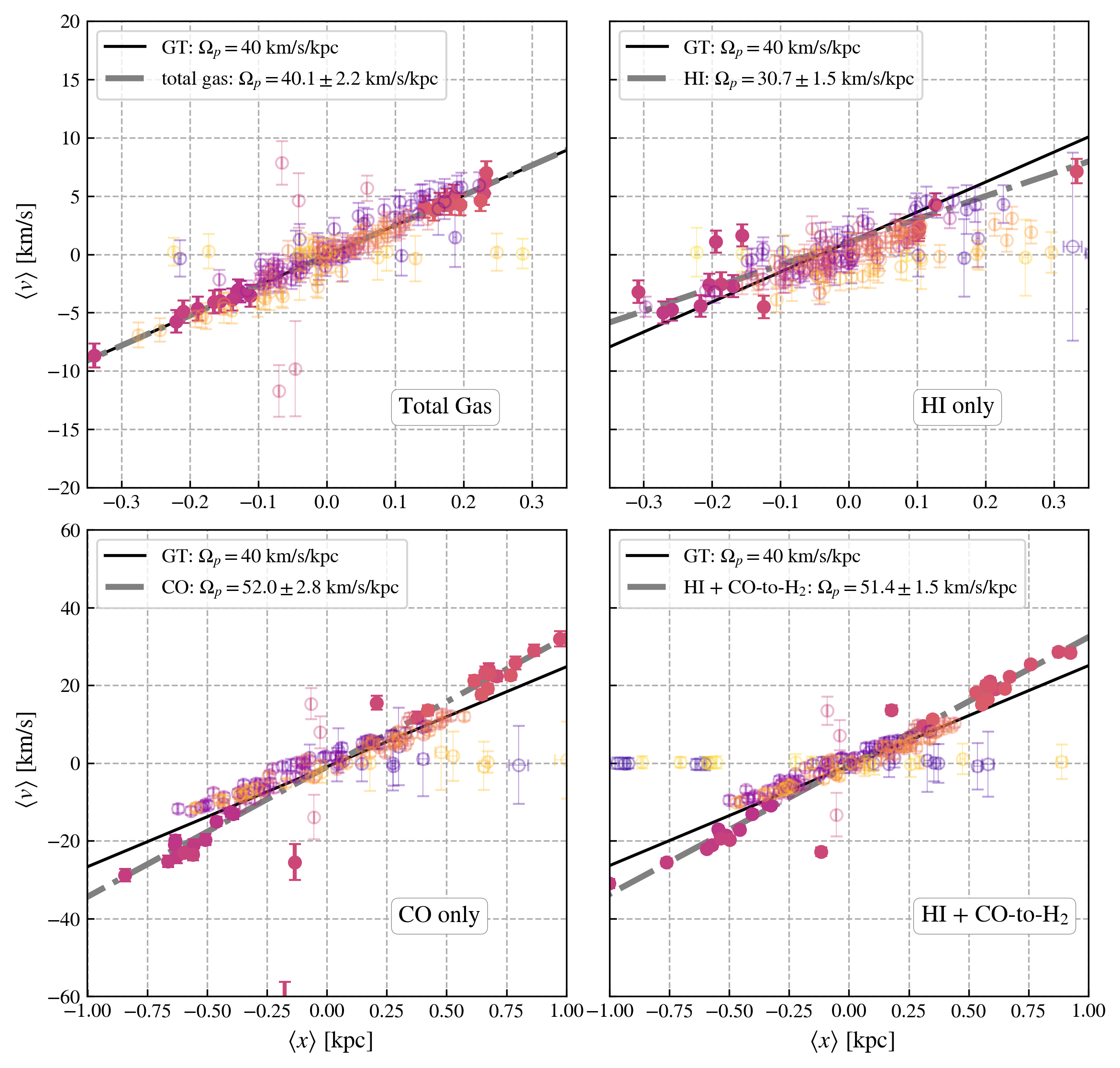}
    \caption{TW method output for our 3D simulation when applied to different gas tracers with $i = 40 ^\circ$ and $\psi_{\text{bar}} = 20^\circ$:
    \textit{Top left:} Total gas density;
    \textit{top right:} H{\sc i} gas tracers;
    \textit{bottom left :} CO density, and;
    \textit{bottom right:} Combination of H{\sc i} and  estimation of H$_2$ obtained from CO, including a conversion factor to mimic how molecular gas masses are obtained observationally.
    The black line shows the ground truth pattern speed and the grey solid line is the fit to slits that cross the bar. These slits are shown as filled circles, and slits which are not included in the fitting are marked with open circles.}
    \label{fig:3D}
\end{figure*}

In Figure \ref{fig:3D}  we show results of applying the TW method to these tracers. We see that for total gas density the TW method measures the correct pattern speed. However, H{\sc i} alone typically underpredicts the pattern speed, whilst CO overpredicts. Critically, even when combining H{\sc i} and CO a correct pattern speed is not recovered, indicating these gas tracers are unsuitable for measuring pattern speeds.
We repeated this study for another snapshot 5 Myr later and the results are consistent.

\begin{figure}
    \centering
    \includegraphics[width=\columnwidth]{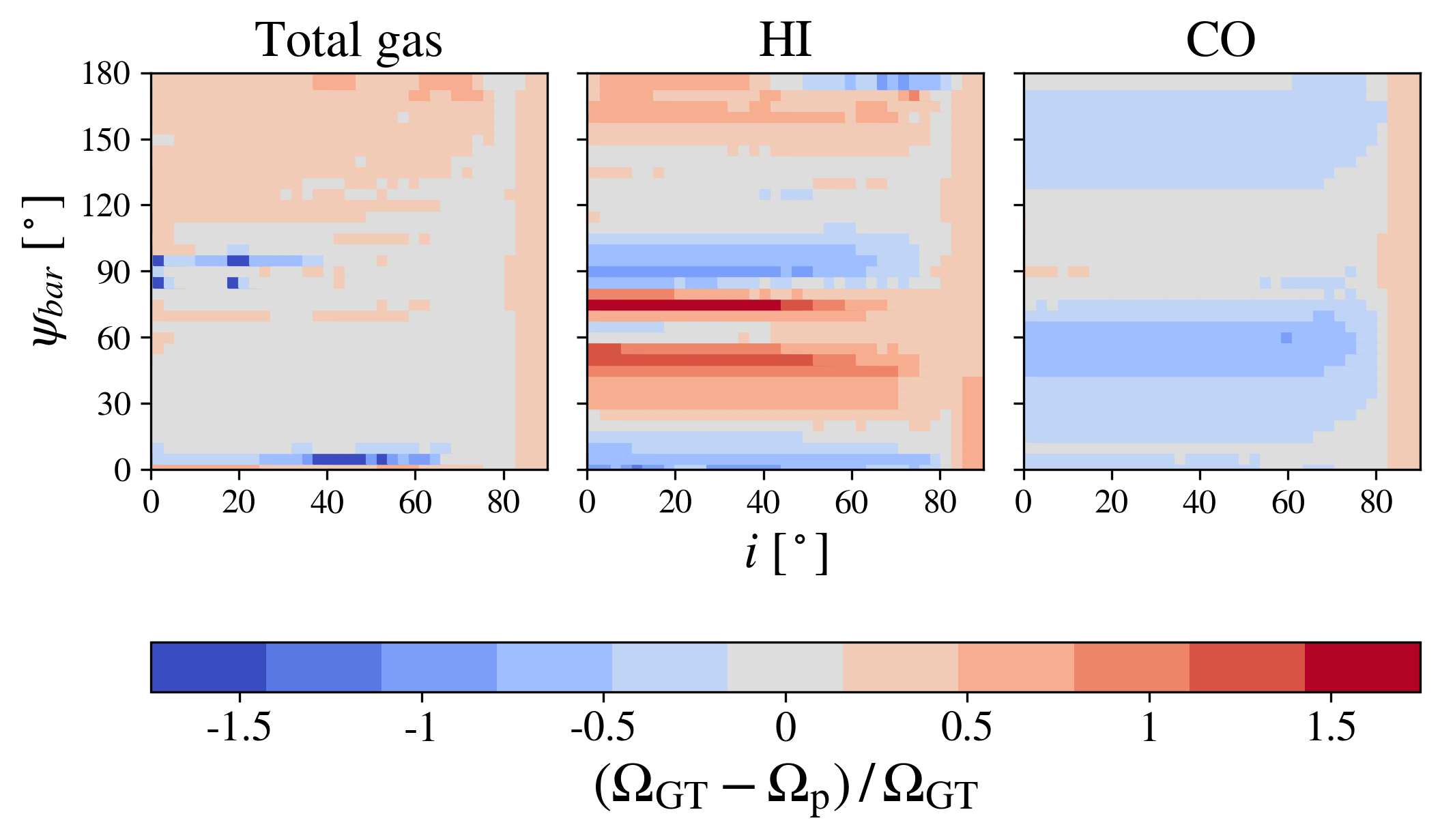}
    \caption{
    The bootstrapped TW method output for a range of angles $i$ and $\psi_{\text{bar}}$ for tracers from 3D simulation. Colour shows the difference between ground truth pattern speed ($\Omega_\text{GT}$) and measured pattern speed ($\Omega_\text{P}$) -- gray indicates agreement and increasing blue (red) shows increasing over- (under-) prediction.
    \textit{Left:} Results for total gas tracer
     \textit{Middle:} HI gas tracer
    \textit{Right:} CO gas tracer.
    }
    \label{fig:bootstrap-3D}
\end{figure}

As illustrated in Figure \ref{fig:bootstrap-3D}, these results are recovered for a range of different geometric configurations (as studied for the 2D models in  sect. \ref{sect:i-psibar-2D}).  The true pattern speed is most reliably extracted using the total gas tracer (where the gas obeys the continuity equation) at the majority of angle combinations (see Fig. \ref{fig:bootstrap-3D}), yielding an overall picture similar to 2D simulation test (see right panel of Fig. \ref{fig:bootstrap}).  On the other hand, the pattern speed is typically either over- or under-estimated using the CO (right panel of Fig. \ref{fig:bootstrap-3D}) or H{\sc i} (middle panel of Fig. \ref{fig:bootstrap-3D}) alone, respectively. It is noteworthy that there is a small range of inclincations $i \in [110 ^\circ; 120 ^\circ]$ where the true pattern speed is recovered by both tracers. We speculate that this is the result of the geometry of the bar; for these angles velocities in the bar region are higher than in the outskirts of the galaxy.

\section{Discussion}
\label{sect:discussion}

Our work shows that for more highly inclined galaxies, the uncertainty limit in the position angle is more strict, which is opposite to the conclusion presented by \citet[see fig. 8]{Debattista2003}. In theory, the more inclined galaxy is, the more asymmetric it becomes when a position angle uncertainty $\delta_{\text{PA}}$ is included, therefore the method should not measure the correct pattern speed for high values of both inclination $i$ and PA uncertainty $\delta_{\text{PA}}$. We have alluded to why this occurs in our Section \ref{sect:perfectdisk} -- at higher inclination, we often catch the angular rotation curve of the galaxy, which can be confused with a pattern speed. It is likely for the more highly inclined simulations in \citet{Debattista2003}, this effect is dominating the ``pattern speed'' measurement, but is incorrectly identified as the pattern speed.

Furthermore, we applied the TW method to 3D simulations to study how well the GT pattern speed can be extracted using different gas phases as independent kinematic tracers. We show that the TW method provides different results depending on which gas phase we use, even though they respond to the same bar potential.
We attribute this difference primarily to the small-scale morphology of the different tracers. CO is very clumpy, which leads to measurements that consist of both pattern speed and average velocity field.
We can similarly conclude that we observe higher ``pattern speed'' in the center of the 2D galaxy (Fig. \ref{fig:strongbar}) due to clumps and holes in this unsteady region, so we sum up high orbital rotation velocity with the pattern speed.

The H{\sc i} density map has different properties, it is much smoother and has less density in the bar relative to the outer galactic disc (see density maps in Appendix \ref{sect:densitymaps}). Therefore, weights in the integrals in Eq. \ref{eq:TWM} corresponding to the bar region are not sufficiently strong and the signal tends to be partially cancelled out by pixels at larger galactocentric radius. 

\cite{Williams2021} showed that for observational data, we get the same result as we have shown here -- that the pattern speed measured by CO is higher than the true pattern speed. Moreover, we suppose that slow bars measured in dwarf galaxies \citep{Bureau1999, Banerjee2013} might be an outcome of the inapplicability of the TW method to H{\sc i}. However, we need to stress that dwarf galaxies are H{\sc i} dominated so the continuity equation may work better for this gas phase there than in our simulation.

Furthermore, gas in our simulation is more idealized than in any observations. First, real galaxies are much less symmetric which is crucial for this method. Second, the simulations have a single well-defined pattern from a rigidly rotation bar, while real galaxies will have also additional perturbations coming from spiral arms, interaction with other galaxies, satellites, etc. 
We also have not studied the effects that star formation has on the gas distribution and kinematics -- given that dwarf galaxies typically have higher star formation rate efficiencies than normal spirals \citep[e.g.][]{2008Leroy}, this effect may be significant. However, investigating the effects of star formation and different galaxy classes is beyond the scope of this work -- we leave this to future efforts.

\section{Conclusions}
\label{sect:conclusions}

In this work, we have applied the Tremaine-Weinberg (TW) method to a number of simple simulations, to test where this method is able to recover correct pattern speeds, and where its reliability is compromised. We have shown that the method is often unsuitable, but our recommendations for applying it are as follows:

Firstly, we would advise that the TW method can be applied only to galaxies with inclinations $i \in [5^{\circ}, 50^{\circ}]$. The lower limit depends on the signal-to-noise ratio for available data, while the upper one is set by the TW method catching other sources of the signal rather than the pattern speed. This is a more strict range than recommended by earlier studies \citep{Guo2019, Zou2019}, where the range was wider and inclinations up to 70$^\circ$ were allowed. However, we want to stress that this conclusion is based on very simple and unrealistic model, and therefore we would suggest studies adapt such a test to match their own data.

Secondly, there is also a limitation for a bar (mis-)alignment angle. Previous studies with $N$-body simulations \citep{Zou2019} showed that the TW method works for a wide range $10^{\circ} < |\psi_{\text{bar}}| < 75^{\circ}$. Although our 2D simulation test (Fig. \ref{fig:bootstrap}) agrees with this conclusion, when we also include extra solid body rotation, it is clear that the range should be narrowed. Thus, the bar should be oriented towards the major axis with a misalignment angle $5^{\circ} < |\psi_{\text{bar}}| < 30^{\circ}$.

Thirdly, the PA should be measured with an error not higher than $1^{\circ}$ for a galaxy inclination of $50^{\circ}$. For less inclined galaxies, the uncertainty in PA can be up to $10^{\circ}$ (see Fig. \ref{fig:deltaPA_incl}).

Finally, by applying the TW method to a 3D simulation we conclude that the method produces incorrect results when applied to gas tracers, due to both a violation of the continuity equation and the gas morphology. Using CO data typically leads to overestimated values of pattern speed, while H{\sc i} data leads to an underestimation.

Overall, this work shows that the TW method should be used with extreme caution and using strict criteria when applied to ISM tracers. Given the overall simplicity of our tests, we expect that these criteria will become even more strict when additional processes such as star formation, a live stellar potential or galaxy interactions are included. Further tests using more sophisticated galaxy simulations will be critical for assessing how well we can measure the pattern speeds of bars in galaxies using the optical IFU instruments that are now regularly producing maps of the stellar surface density and kinematics.

\section*{Acknowledgements}

This work has been carried out during a Summer Internship funded by the Max Planck Society.

%%%%%%%%%%%%%%%%%%%%%%%%%%%%%%%%%%%%%%%%%%%%%%%%%%
\section*{Data Availability}

The simulation's snapshots and interactive plots are available on \href{https://github.com/olgaborodina/TWM_public}{GitHub} 



\bibliographystyle{mnras}
\bibliography{extragal} % if your bibtex file is called example.bib

\begin{thebibliography}{}
\makeatletter
\relax
\def\mn@urlcharsother{\let\do\@makeother \do\$\do\&\do\#\do\^\do\_\do\%\do\~}
\def\mn@doi{\begingroup\mn@urlcharsother \@ifnextchar [ {\mn@doi@}
  {\mn@doi@[]}}
\def\mn@doi@[#1]#2{\def\@tempa{#1}\ifx\@tempa\@empty \href
  {http://dx.doi.org/#2} {doi:#2}\else \href {http://dx.doi.org/#2} {#1}\fi
  \endgroup}
\def\mn@eprint#1#2{\mn@eprint@#1:#2::\@nil}
\def\mn@eprint@arXiv#1{\href {http://arxiv.org/abs/#1} {{\tt arXiv:#1}}}
\def\mn@eprint@dblp#1{\href {http://dblp.uni-trier.de/rec/bibtex/#1.xml}
  {dblp:#1}}
\def\mn@eprint@#1:#2:#3:#4\@nil{\def\@tempa {#1}\def\@tempb {#2}\def\@tempc
  {#3}\ifx \@tempc \@empty \let \@tempc \@tempb \let \@tempb \@tempa \fi \ifx
  \@tempb \@empty \def\@tempb {arXiv}\fi \@ifundefined
  {mn@eprint@\@tempb}{\@tempb:\@tempc}{\expandafter \expandafter \csname
  mn@eprint@\@tempb\endcsname \expandafter{\@tempc}}}

\bibitem[\protect\citeauthoryear{{Aguerri}, {M{\'e}ndez-Abreu}  \&
  {Corsini}}{{Aguerri} et~al.}{2009}]{Aguerri2009}
{Aguerri} J.~A.~L.,  {M{\'e}ndez-Abreu} J.,   {Corsini} E.~M.,  2009, \mn@doi
  [\aap] {10.1051/0004-6361:200810931}, \href
  {https://ui.adsabs.harvard.edu/abs/2009A&A...495..491A} {495, 491}

\bibitem[\protect\citeauthoryear{{Athanassoula}}{{Athanassoula}}{1992}]{Athanassoula1992}
{Athanassoula} E.,  1992, \mn@doi [\mnras] {10.1093/mnras/259.2.345}, \href
  {https://ui.adsabs.harvard.edu/abs/1992MNRAS.259..345A} {259, 345}

\bibitem[\protect\citeauthoryear{{Banerjee}, {Patra}, {Chengalur}  \&
  {Begum}}{{Banerjee} et~al.}{2013}]{Banerjee2013}
{Banerjee} A.,  {Patra} N.~N.,  {Chengalur} J.~N.,   {Begum} A.,  2013, \mn@doi
  [\mnras] {10.1093/mnras/stt1083}, \href
  {https://ui.adsabs.harvard.edu/abs/2013MNRAS.434.1257B} {434, 1257}

\bibitem[\protect\citeauthoryear{{Beane} et~al.,}{{Beane}
  et~al.}{2022}]{Beane2022}
{Beane} A.,  et~al., 2022, arXiv e-prints, \href
  {https://ui.adsabs.harvard.edu/abs/2022arXiv220903364B} {p. arXiv:2209.03364}

\bibitem[\protect\citeauthoryear{{Beuther}, {Meidt}, {Schinnerer}, {Paladino}
  \& {Leroy}}{{Beuther} et~al.}{2017}]{Beuther2017}
{Beuther} H.,  {Meidt} S.,  {Schinnerer} E.,  {Paladino} R.,   {Leroy} A.,
  2017, \mn@doi [\aap] {10.1051/0004-6361/201526749}, \href
  {https://ui.adsabs.harvard.edu/abs/2017A&A...597A..85B} {597, A85}

\bibitem[\protect\citeauthoryear{{Binney} \& {Tremaine}}{{Binney} \&
  {Tremaine}}{2008}]{BinneyTremaine2008}
{Binney} J.,  {Tremaine} S.,  2008, {Galactic Dynamics: Second Edition}

\bibitem[\protect\citeauthoryear{{Bureau}, {Freeman}, {Pfitzner}  \&
  {Meurer}}{{Bureau} et~al.}{1999}]{Bureau1999}
{Bureau} M.,  {Freeman} K.~C.,  {Pfitzner} D.~W.,   {Meurer} G.~R.,  1999,
  \mn@doi [\aj] {10.1086/301064}, \href
  {https://ui.adsabs.harvard.edu/abs/1999AJ....118.2158B} {118, 2158}

\bibitem[\protect\citeauthoryear{{Buttitta} et~al.,}{{Buttitta}
  et~al.}{2022}]{Buttitta2022}
{Buttitta} C.,  et~al., 2022, \mn@doi [\aap] {10.1051/0004-6361/202244297},
  \href {https://ui.adsabs.harvard.edu/abs/2022A&A...664L..10B} {664, L10}

\bibitem[\protect\citeauthoryear{{Chemin} \& {Hernandez}}{{Chemin} \&
  {Hernandez}}{2009}]{Chemin2009}
{Chemin} L.,  {Hernandez} O.,  2009, \mn@doi [\aap]
  {10.1051/0004-6361/200912019}, \href
  {https://ui.adsabs.harvard.edu/abs/2009A&A...499L..25C} {499, L25}

\bibitem[\protect\citeauthoryear{{Corsini}, {Aguerri}, {Debattista},
  {Pizzella}, {Barazza}  \& {Jerjen}}{{Corsini} et~al.}{2007}]{Corsini2007}
{Corsini} E.~M.,  {Aguerri} J.~A.~L.,  {Debattista} V.~P.,  {Pizzella} A.,
  {Barazza} F.~D.,   {Jerjen} H.,  2007, \mn@doi [\apjl] {10.1086/518035},
  \href {https://ui.adsabs.harvard.edu/abs/2007ApJ...659L.121C} {659, L121}

\bibitem[\protect\citeauthoryear{{Cuomo}, {Lopez Aguerri}, {Corsini},
  {Debattista}, {M{\'e}ndez-Abreu}  \& {Pizzella}}{{Cuomo}
  et~al.}{2019}]{Cuomo2019}
{Cuomo} V.,  {Lopez Aguerri} J.~A.,  {Corsini} E.~M.,  {Debattista} V.~P.,
  {M{\'e}ndez-Abreu} J.,   {Pizzella} A.,  2019, \mn@doi [\aap]
  {10.1051/0004-6361/201936415}, \href
  {https://ui.adsabs.harvard.edu/abs/2019A&A...632A..51C} {632, A51}

\bibitem[\protect\citeauthoryear{{Cuomo}, {Aguerri}, {Corsini}  \&
  {Debattista}}{{Cuomo} et~al.}{2020}]{Cuomo2020}
{Cuomo} V.,  {Aguerri} J.~A.~L.,  {Corsini} E.~M.,   {Debattista} V.~P.,  2020,
  \mn@doi [\aap] {10.1051/0004-6361/202037945}, \href
  {https://ui.adsabs.harvard.edu/abs/2020A&A...641A.111C} {641, A111}

\bibitem[\protect\citeauthoryear{{Debattista}}{{Debattista}}{2003}]{Debattista2003}
{Debattista} V.~P.,  2003, \mn@doi [\mnras] {10.1046/j.1365-8711.2003.06620.x},
  \href {https://ui.adsabs.harvard.edu/abs/2003MNRAS.342.1194D} {342, 1194}

\bibitem[\protect\citeauthoryear{{Debattista} \& {Sellwood}}{{Debattista} \&
  {Sellwood}}{2000}]{Debattista2000}
{Debattista} V.~P.,  {Sellwood} J.~A.,  2000, \mn@doi [\apj] {10.1086/317148},
  \href {https://ui.adsabs.harvard.edu/abs/2000ApJ...543..704D} {543, 704}

\bibitem[\protect\citeauthoryear{{Debattista} \& {Williams}}{{Debattista} \&
  {Williams}}{2004}]{Debattista2004}
{Debattista} V.~P.,  {Williams} T.~B.,  2004, \mn@doi [\apj] {10.1086/382585},
  \href {https://ui.adsabs.harvard.edu/abs/2004ApJ...605..714D} {605, 714}

\bibitem[\protect\citeauthoryear{{Elmegreen}, {Elmegreen}  \&
  {Seiden}}{{Elmegreen} et~al.}{1989}]{Elmegreen1989}
{Elmegreen} B.~G.,  {Elmegreen} D.~M.,   {Seiden} P.~E.,  1989, \mn@doi [\apj]
  {10.1086/167733}, \href
  {https://ui.adsabs.harvard.edu/abs/1989ApJ...343..602E} {343, 602}

\bibitem[\protect\citeauthoryear{{Garma-Oehmichen}, {Cano-D{\'\i}az},
  {Hern{\'a}ndez-Toledo}, {Aquino-Ort{\'\i}z}, {Valenzuela}, {Aguerri},
  {S{\'a}nchez}  \& {Merrifield}}{{Garma-Oehmichen}
  et~al.}{2020}]{GarmaOehmichen2020}
{Garma-Oehmichen} L.,  {Cano-D{\'\i}az} M.,  {Hern{\'a}ndez-Toledo} H.,
  {Aquino-Ort{\'\i}z} E.,  {Valenzuela} O.,  {Aguerri} J.~A.~L.,  {S{\'a}nchez}
  S.~F.,   {Merrifield} M.,  2020, \mn@doi [\mnras] {10.1093/mnras/stz3101},
  \href {https://ui.adsabs.harvard.edu/abs/2020MNRAS.491.3655G} {491, 3655}

\bibitem[\protect\citeauthoryear{{Gerssen} \& {Debattista}}{{Gerssen} \&
  {Debattista}}{2007}]{Gerssen2007}
{Gerssen} J.,  {Debattista} V.~P.,  2007, \mn@doi [\mnras]
  {10.1111/j.1365-2966.2007.11761.x}, \href
  {https://ui.adsabs.harvard.edu/abs/2007MNRAS.378..189G} {378, 189}

\bibitem[\protect\citeauthoryear{{Gerssen}, {Kuijken}  \&
  {Merrifield}}{{Gerssen} et~al.}{2003}]{Gerssen2003}
{Gerssen} J.,  {Kuijken} K.,   {Merrifield} M.~R.,  2003, \mn@doi [\mnras]
  {10.1046/j.1365-8711.2003.06937.x}, \href
  {https://ui.adsabs.harvard.edu/abs/2003MNRAS.345..261G} {345, 261}

\bibitem[\protect\citeauthoryear{{Guo}, {Mao}, {Athanassoula}, {Li}, {Ge},
  {Long}, {Merrifield}  \& {Masters}}{{Guo} et~al.}{2019}]{Guo2019}
{Guo} R.,  {Mao} S.,  {Athanassoula} E.,  {Li} H.,  {Ge} J.,  {Long} R.~J.,
  {Merrifield} M.,   {Masters} K.,  2019, \mn@doi [\mnras]
  {10.1093/mnras/sty2715}, \href
  {https://ui.adsabs.harvard.edu/abs/2019MNRAS.482.1733G} {482, 1733}

\bibitem[\protect\citeauthoryear{{Hernquist} \& {Weinberg}}{{Hernquist} \&
  {Weinberg}}{1992}]{Hernquist1992}
{Hernquist} L.,  {Weinberg} M.~D.,  1992, \mn@doi [\apj] {10.1086/171975},
  \href {https://ui.adsabs.harvard.edu/abs/1992ApJ...400...80H} {400, 80}

\bibitem[\protect\citeauthoryear{{Hirota}, {Kuno}, {Sato}, {Nakanishi},
  {Tosaki}, {Matsui}, {Habe}  \& {Sorai}}{{Hirota} et~al.}{2009}]{Hirota2009}
{Hirota} A.,  {Kuno} N.,  {Sato} N.,  {Nakanishi} H.,  {Tosaki} T.,  {Matsui}
  H.,  {Habe} A.,   {Sorai} K.,  2009, \mn@doi [\pasj] {10.1093/pasj/61.3.441},
  \href {https://ui.adsabs.harvard.edu/abs/2009PASJ...61..441H} {61, 441}

\bibitem[\protect\citeauthoryear{{Hunter}, {Ball}, {Huntley}, {England}  \&
  {Gottesman}}{{Hunter} et~al.}{1988}]{Hunter1988}
{Hunter} J.~H. J.,  {Ball} R.,  {Huntley} J.~M.,  {England} M.~N.,
  {Gottesman} S.~T.,  1988, \mn@doi [\apj] {10.1086/165932}, \href
  {https://ui.adsabs.harvard.edu/abs/1988ApJ...324..721H} {324, 721}

\bibitem[\protect\citeauthoryear{{Kalapotharakos}, {Patsis}  \&
  {Grosb{\o}l}}{{Kalapotharakos} et~al.}{2010}]{Kalapotharakos2010}
{Kalapotharakos} C.,  {Patsis} P.~A.,   {Grosb{\o}l} P.,  2010, \mn@doi
  [\mnras] {10.1111/j.1365-2966.2010.17061.x}, \href
  {https://ui.adsabs.harvard.edu/abs/2010MNRAS.408....9K} {408, 9}

\bibitem[\protect\citeauthoryear{{Kranz}, {Slyz}  \& {Rix}}{{Kranz}
  et~al.}{2003}]{Kranz2003}
{Kranz} T.,  {Slyz} A.,   {Rix} H.-W.,  2003, \mn@doi [\apj] {10.1086/367551},
  \href {https://ui.adsabs.harvard.edu/abs/2003ApJ...586..143K} {586, 143}

\bibitem[\protect\citeauthoryear{{Kuno}, {Nishiyama}, {Nakai}, {Sorai},
  {Vila-Vilar{\'o}}  \& {Handa}}{{Kuno} et~al.}{2000}]{Kuno2000}
{Kuno} N.,  {Nishiyama} K.,  {Nakai} N.,  {Sorai} K.,  {Vila-Vilar{\'o}} B.,
  {Handa} T.,  2000, \mn@doi [\pasj] {10.1093/pasj/52.5.775}, \href
  {https://ui.adsabs.harvard.edu/abs/2000PASJ...52..775K} {52, 775}

\bibitem[\protect\citeauthoryear{{Leroy}, {Walter}, {Brinks}, {Bigiel}, {de
  Blok}, {Madore}  \& {Thornley}}{{Leroy} et~al.}{2008}]{2008Leroy}
{Leroy} A.~K.,  {Walter} F.,  {Brinks} E.,  {Bigiel} F.,  {de Blok} W.~J.~G.,
  {Madore} B.,   {Thornley} M.~D.,  2008, \mn@doi [\aj]
  {10.1088/0004-6256/136/6/2782}, \href
  {https://ui.adsabs.harvard.edu/abs/2008AJ....136.2782L} {136, 2782}

\bibitem[\protect\citeauthoryear{{Leroy} et~al.,}{{Leroy}
  et~al.}{2021}]{Leroy2021}
{Leroy} A.~K.,  et~al., 2021, \mn@doi [\apjs] {10.3847/1538-4365/ac17f3}, \href
  {https://ui.adsabs.harvard.edu/abs/2021ApJS..257...43L} {257, 43}

\bibitem[\protect\citeauthoryear{{Lin}, {Wang}, {Hsieh}, {Taam}, {Yang}  \&
  {Yen}}{{Lin} et~al.}{2013}]{Lin2013}
{Lin} L.-H.,  {Wang} H.-H.,  {Hsieh} P.-Y.,  {Taam} R.~E.,  {Yang} C.-C.,
  {Yen} D. C.~C.,  2013, \mn@doi [\apj] {10.1088/0004-637X/771/1/8}, \href
  {https://ui.adsabs.harvard.edu/abs/2013ApJ...771....8L} {771, 8}

\bibitem[\protect\citeauthoryear{{Mignone}, {Bodo}, {Massaglia}, {Matsakos},
  {Tesileanu}, {Zanni}  \& {Ferrari}}{{Mignone} et~al.}{2007}]{Mignone2007}
{Mignone} A.,  {Bodo} G.,  {Massaglia} S.,  {Matsakos} T.,  {Tesileanu} O.,
  {Zanni} C.,   {Ferrari} A.,  2007, \mn@doi [\apjs] {10.1086/513316}, \href
  {https://ui.adsabs.harvard.edu/abs/2007ApJS..170..228M} {170, 228}

\bibitem[\protect\citeauthoryear{{Nelson} et~al.,}{{Nelson}
  et~al.}{2018}]{Nelson2018}
{Nelson} D.,  et~al., 2018, \mn@doi [\mnras] {10.1093/mnras/stx3040}, \href
  {https://ui.adsabs.harvard.edu/abs/2018MNRAS.475..624N} {475, 624}

\bibitem[\protect\citeauthoryear{{Pillepich} et~al.,}{{Pillepich}
  et~al.}{2019}]{Pillepich2019}
{Pillepich} A.,  et~al., 2019, \mn@doi [\mnras] {10.1093/mnras/stz2338}, \href
  {https://ui.adsabs.harvard.edu/abs/2019MNRAS.490.3196P} {490, 3196}

\bibitem[\protect\citeauthoryear{{Querejeta} et~al.,}{{Querejeta}
  et~al.}{2021}]{Querejeta2021}
{Querejeta} M.,  et~al., 2021, \mn@doi [\aap] {10.1051/0004-6361/202140695},
  \href {https://ui.adsabs.harvard.edu/abs/2021A&A...656A.133Q} {656, A133}

\bibitem[\protect\citeauthoryear{{Rautiainen}, {Salo}  \&
  {Laurikainen}}{{Rautiainen} et~al.}{2008}]{Rautiainen2008}
{Rautiainen} P.,  {Salo} H.,   {Laurikainen} E.,  2008, \mn@doi [\mnras]
  {10.1111/j.1365-2966.2008.13522.x}, \href
  {https://ui.adsabs.harvard.edu/abs/2008MNRAS.388.1803R} {388, 1803}

\bibitem[\protect\citeauthoryear{{Ridley}, {Sormani}, {Tre{\ss}}, {Magorrian}
  \& {Klessen}}{{Ridley} et~al.}{2017}]{Ridley2017}
{Ridley} M. G.~L.,  {Sormani} M.~C.,  {Tre{\ss}} R.~G.,  {Magorrian} J.,
  {Klessen} R.~S.,  2017, \mn@doi [\mnras] {10.1093/mnras/stx944}, \href
  {https://ui.adsabs.harvard.edu/abs/2017MNRAS.469.2251R} {469, 2251}

\bibitem[\protect\citeauthoryear{{Roshan}, {Ghafourian}, {Kashfi}, {Banik},
  {Haslbauer}, {Cuomo}, {Famaey}  \& {Kroupa}}{{Roshan}
  et~al.}{2021}]{Roshan2021}
{Roshan} M.,  {Ghafourian} N.,  {Kashfi} T.,  {Banik} I.,  {Haslbauer} M.,
  {Cuomo} V.,  {Famaey} B.,   {Kroupa} P.,  2021, \mn@doi [\mnras]
  {10.1093/mnras/stab2553}, \href
  {https://ui.adsabs.harvard.edu/abs/2021MNRAS.508..926R} {508, 926}

\bibitem[\protect\citeauthoryear{{Salo}, {Rautiainen}, {Buta}, {Purcell},
  {Cobb}, {Crocker}  \& {Laurikainen}}{{Salo} et~al.}{1999}]{Salo1999}
{Salo} H.,  {Rautiainen} P.,  {Buta} R.,  {Purcell} G.~B.,  {Cobb} M.~L.,
  {Crocker} D.~A.,   {Laurikainen} E.,  1999, \mn@doi [\aj] {10.1086/300726},
  \href {https://ui.adsabs.harvard.edu/abs/1999AJ....117..792S} {117, 792}

\bibitem[\protect\citeauthoryear{{Schaye} et~al.,}{{Schaye}
  et~al.}{2015}]{Schaye2015}
{Schaye} J.,  et~al., 2015, \mn@doi [\mnras] {10.1093/mnras/stu2058}, \href
  {https://ui.adsabs.harvard.edu/abs/2015MNRAS.446..521S} {446, 521}

\bibitem[\protect\citeauthoryear{{Schinnerer} et~al.,}{{Schinnerer}
  et~al.}{2019a}]{Schinnerer2019Msngr}
{Schinnerer} E.,  et~al., 2019a, \mn@doi [The Messenger]
  {10.18727/0722-6691/5151}, \href
  {https://ui.adsabs.harvard.edu/abs/2019Msngr.177...36S} {177, 36}

\bibitem[\protect\citeauthoryear{{Schinnerer} et~al.,}{{Schinnerer}
  et~al.}{2019b}]{Schinnerer2019}
{Schinnerer} E.,  et~al., 2019b, \mn@doi [\apj] {10.3847/1538-4357/ab50c2},
  \href {https://ui.adsabs.harvard.edu/abs/2019ApJ...887...49S} {887, 49}

\bibitem[\protect\citeauthoryear{{Sellwood} \& {Wilkinson}}{{Sellwood} \&
  {Wilkinson}}{1993}]{Sellwood1993}
{Sellwood} J.~A.,  {Wilkinson} A.,  1993, \mn@doi [Reports on Progress in
  Physics] {10.1088/0034-4885/56/2/001}, \href
  {https://ui.adsabs.harvard.edu/abs/1993RPPh...56..173S} {56, 173}

\bibitem[\protect\citeauthoryear{{Sempere}, {Garcia-Burillo}, {Combes}  \&
  {Knapen}}{{Sempere} et~al.}{1995}]{Sempere1995}
{Sempere} M.~J.,  {Garcia-Burillo} S.,  {Combes} F.,   {Knapen} J.~H.,  1995,
  \aap, \href {https://ui.adsabs.harvard.edu/abs/1995A&A...296...45S} {296, 45}

\bibitem[\protect\citeauthoryear{{Sheth} et~al.,}{{Sheth}
  et~al.}{2008}]{Sheth2008}
{Sheth} K.,  et~al., 2008, \mn@doi [\apj] {10.1086/524980}, \href
  {https://ui.adsabs.harvard.edu/abs/2008ApJ...675.1141S} {675, 1141}

\bibitem[\protect\citeauthoryear{{Sormani}, {Binney}  \& {Magorrian}}{{Sormani}
  et~al.}{2015}]{Sormani2015}
{Sormani} M.~C.,  {Binney} J.,   {Magorrian} J.,  2015, \mn@doi [\mnras]
  {10.1093/mnras/stv2067}, \href
  {https://ui.adsabs.harvard.edu/abs/2015MNRAS.454.1818S} {454, 1818}

\bibitem[\protect\citeauthoryear{{Sormani}, {Tre{\ss}}, {Ridley}, {Glover},
  {Klessen}, {Binney}, {Magorrian}  \& {Smith}}{{Sormani}
  et~al.}{2018}]{Sormani2018}
{Sormani} M.~C.,  {Tre{\ss}} R.~G.,  {Ridley} M.,  {Glover} S. C.~O.,
  {Klessen} R.~S.,  {Binney} J.,  {Magorrian} J.,   {Smith} R.,  2018, \mn@doi
  [\mnras] {10.1093/mnras/stx3258}, \href
  {https://ui.adsabs.harvard.edu/abs/2018MNRAS.475.2383S} {475, 2383}

\bibitem[\protect\citeauthoryear{{Springel}}{{Springel}}{2010}]{Springel2010}
{Springel} V.,  2010, \mn@doi [\mnras] {10.1111/j.1365-2966.2009.15715.x},
  \href {https://ui.adsabs.harvard.edu/abs/2010MNRAS.401..791S} {401, 791}

\bibitem[\protect\citeauthoryear{{Tremaine} \& {Weinberg}}{{Tremaine} \&
  {Weinberg}}{1984}]{TW1984}
{Tremaine} S.,  {Weinberg} M.~D.,  1984, \mn@doi [\apjl] {10.1086/184292},
  \href {https://ui.adsabs.harvard.edu/abs/1984ApJ...282L...5T} {282, L5}

\bibitem[\protect\citeauthoryear{{Virtanen} et~al.,}{{Virtanen}
  et~al.}{2020}]{scipy2020}
{Virtanen} P.,  et~al., 2020, \mn@doi [Nature Methods]
  {10.1038/s41592-019-0686-2}, \href
  {https://ui.adsabs.harvard.edu/abs/2020NatMe..17..261V} {17, 261}

\bibitem[\protect\citeauthoryear{{Weinberg} \& {Katz}}{{Weinberg} \&
  {Katz}}{2007}]{Weinberg2007}
{Weinberg} M.~D.,  {Katz} N.,  2007, \mn@doi [\mnras]
  {10.1111/j.1365-2966.2006.11307.x}, \href
  {https://ui.adsabs.harvard.edu/abs/2007MNRAS.375..460W} {375, 460}

\bibitem[\protect\citeauthoryear{{Weinberger}, {Springel}  \&
  {Pakmor}}{{Weinberger} et~al.}{2020}]{Weinberger2020}
{Weinberger} R.,  {Springel} V.,   {Pakmor} R.,  2020, \mn@doi [\apjs]
  {10.3847/1538-4365/ab908c}, \href
  {https://ui.adsabs.harvard.edu/abs/2020ApJS..248...32W} {248, 32}

\bibitem[\protect\citeauthoryear{{Weiner}, {Sellwood}  \& {Williams}}{{Weiner}
  et~al.}{2001}]{Weiner2001}
{Weiner} B.~J.,  {Sellwood} J.~A.,   {Williams} T.~B.,  2001, \mn@doi [\apj]
  {10.1086/318289}, \href
  {https://ui.adsabs.harvard.edu/abs/2001ApJ...546..931W} {546, 931}

\bibitem[\protect\citeauthoryear{{Williams} et~al.,}{{Williams}
  et~al.}{2021}]{Williams2021}
{Williams} T.~G.,  et~al., 2021, \mn@doi [\aj] {10.3847/1538-3881/abe243},
  \href {https://ui.adsabs.harvard.edu/abs/2021AJ....161..185W} {161, 185}

\bibitem[\protect\citeauthoryear{{Zimmer}, {Rand}  \& {McGraw}}{{Zimmer}
  et~al.}{2004}]{Zimmer2004}
{Zimmer} P.,  {Rand} R.~J.,   {McGraw} J.~T.,  2004, \mn@doi [\apj]
  {10.1086/383459}, \href
  {https://ui.adsabs.harvard.edu/abs/2004ApJ...607..285Z} {607, 285}

\bibitem[\protect\citeauthoryear{{Zou}, {Shen}, {Bureau}  \& {Li}}{{Zou}
  et~al.}{2019}]{Zou2019}
{Zou} Y.,  {Shen} J.,  {Bureau} M.,   {Li} Z.-Y.,  2019, \mn@doi [\apj]
  {10.3847/1538-4357/ab3f34}, \href
  {https://ui.adsabs.harvard.edu/abs/2019ApJ...884...23Z} {884, 23}

\makeatother
\end{thebibliography}

%\begin{thebibliography}{99}
%\end{thebibliography}

%%%%%%%%%%%%%%%%%%%%%%%%%%%%%%%%%%%%%%%%%%%%%%%%%%

%%%%%%%%%%%%%%%%% APPENDICES %%%%%%%%%%%%%%%%%%%%%

\appendix

\section{Density maps}
\label{sect:densitymaps}

\begin{figure}
    \centering
    \includegraphics[width=\columnwidth]{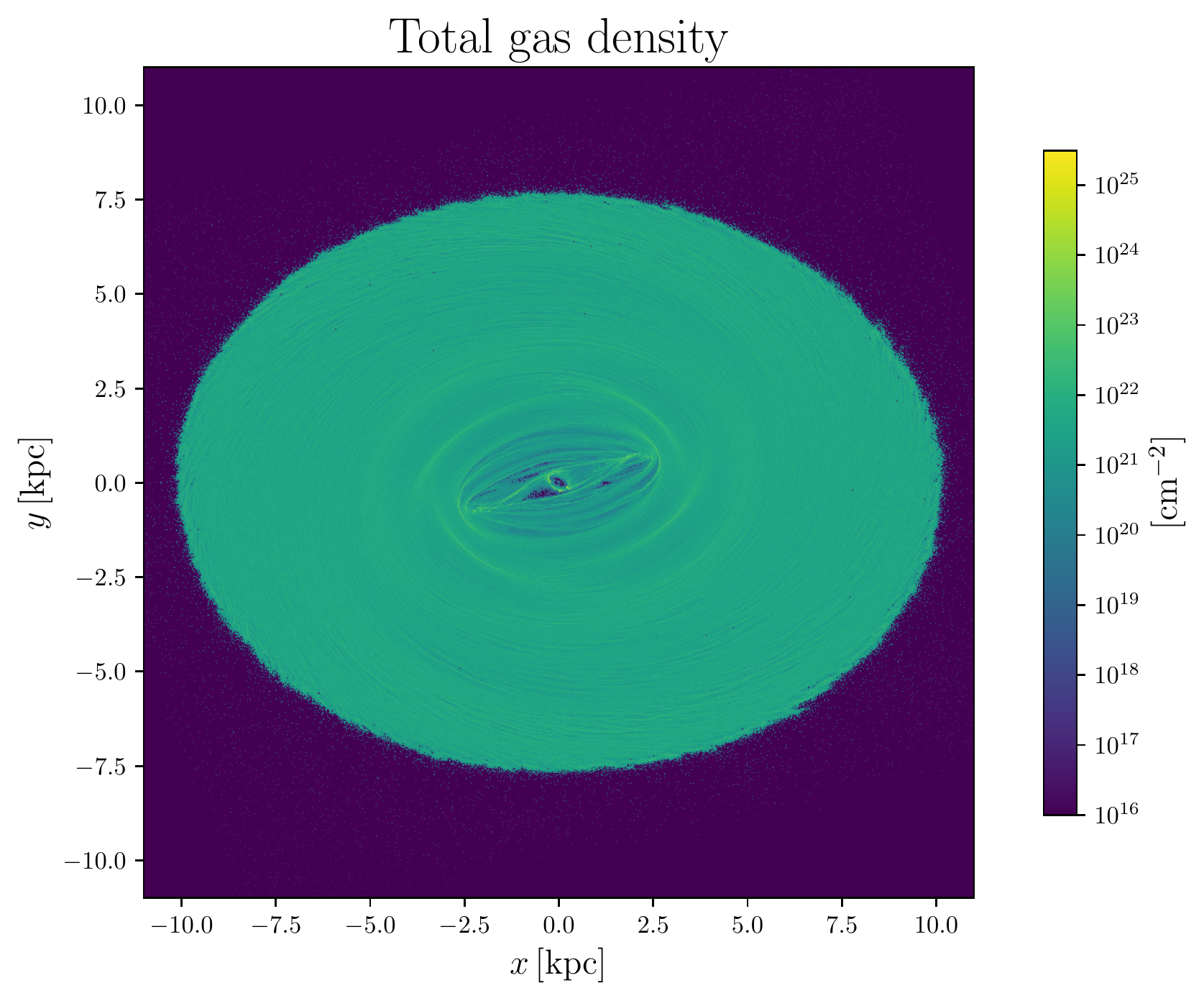}
    \includegraphics[width=\columnwidth]{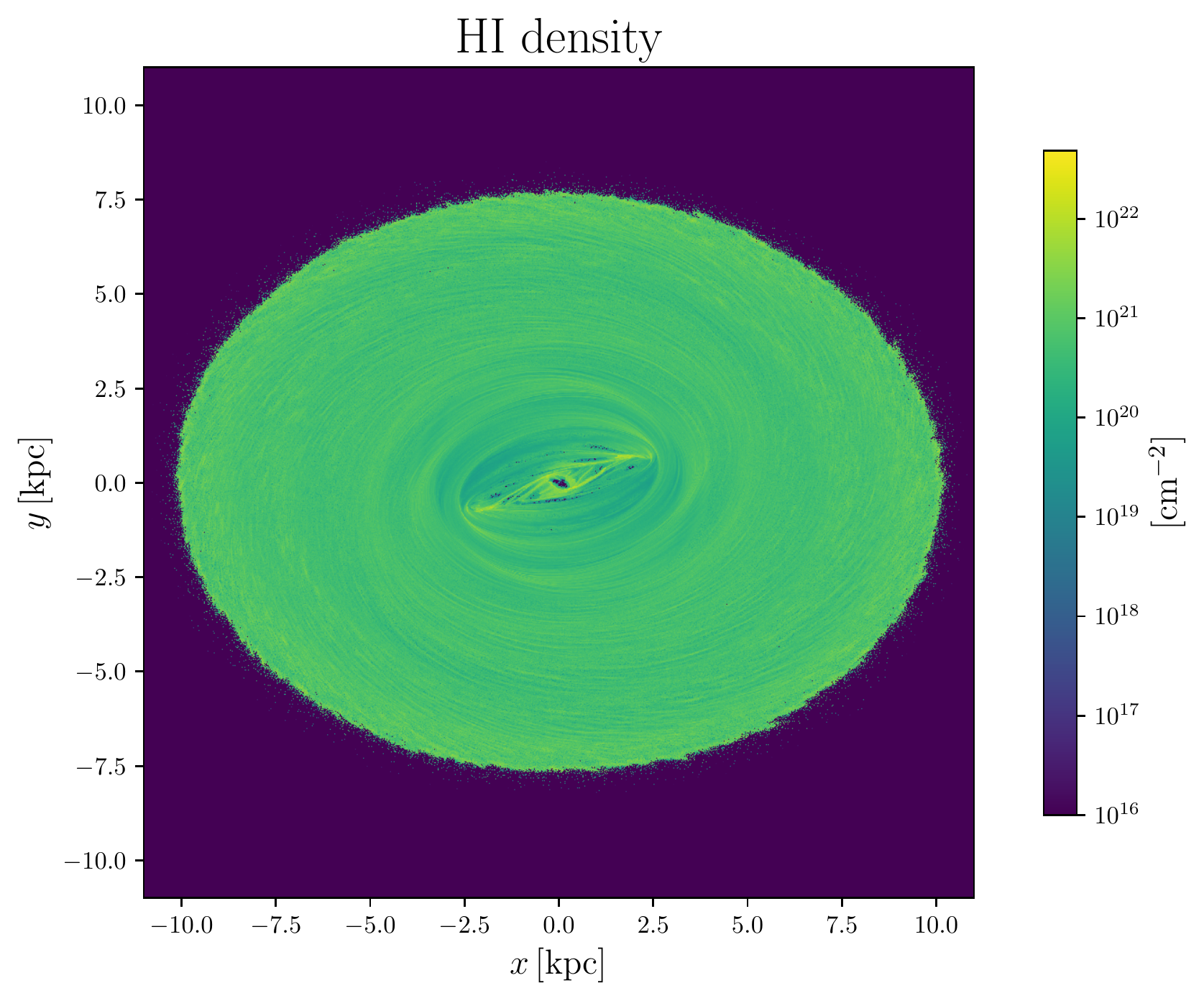}
    \includegraphics[width=\columnwidth]{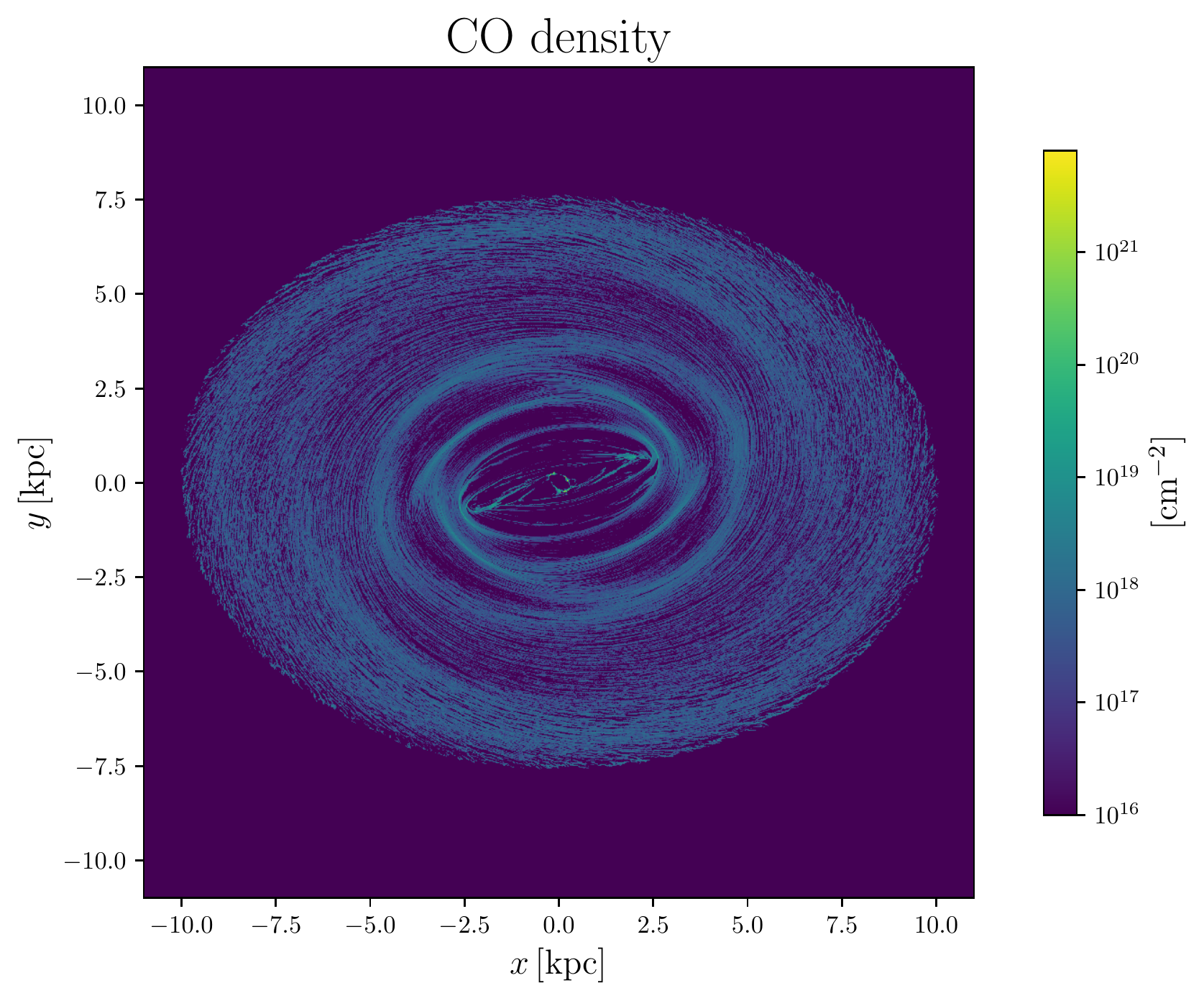}
    \caption{Surface density maps for total gas in the simulation (top panel), H{\sc i} (middle panel) and CO (bottom panel). H{\sc i} is distributed more uniformly and smoothly while CO is very clumpy. The inclination of a galaxy is $50^{\circ}$}
    \label{fig:density_maps}
\end{figure}

Using 3D hydrodynamical simulations \citep{Sormani2018} we applied the TW method to different gas tracers. Here, we show the density maps for total gas, for CO and H{\sc i} (Fig. \ref{fig:density_maps}). Our tests show that the galactic disc should only have asymmetry in the bar region, so that the TW method measures the correct pattern speed. However, the CO density map shows many clumps and holes which leads to the TW method picking up additional signal besides the real pattern speed. For H{\sc i} we see the opposite: the bar region is much less visible and we tend to underestimate the pattern speed.

% If you want to present additional material which would interrupt the flow of the main paper,
% it can be placed in an Appendix which appears after the list of references.

%%%%%%%%%%%%%%%%%%%%%%%%%%%%%%%%%%%%%%%%%%%%%%%%%%

% Don't change these lines
\bsp	% typesetting comment
\label{lastpage}
\end{document}